\documentclass[aps,twocolumn,prd,superscriptaddress]{revtex4-1}

\usepackage[utf8]{inputenc}

\usepackage{mathtools}
\usepackage{amsfonts}
\usepackage{mathrsfs}
\usepackage{bbm}
\usepackage{physics}
\usepackage{slashed}
\usepackage{tensor}

\usepackage{graphicx}
\usepackage{color, float}
\usepackage{array}
\usepackage[abs]{overpic}

\usepackage{placeins}

\usepackage{makecell}
\usepackage{subcaption}

\usepackage{xspace}
\usepackage{siunitx}
\usepackage{xfrac}
\usepackage{hyperref}
\usepackage[nameinlink]{cleveref}
\usepackage{appendix}

\newcommand*{\ie}{i.e.\@\xspace}

\newcommand{\imag}{\text{i}}
\usepackage{xifthen}
\usepackage{xcolor}
\hypersetup{
	colorlinks,
	linkcolor={blue!75!black},
	citecolor={blue!75!black},
	urlcolor={blue!75!black}
}

\usepackage{booktabs}
\usepackage{multirow}

\newcolumntype{C}{>{$}c<{$}}
\AtBeginDocument{
	\heavyrulewidth=.08em
	\lightrulewidth=.05em
	\cmidrulewidth=.03em
	\belowrulesep=.65ex
	\belowbottomsep=0pt
	\aboverulesep=.4ex
	\abovetopsep=0pt
	\cmidrulesep=\doublerulesep
	\cmidrulekern=.5em
	\defaultaddspace=.5em
}

\graphicspath{{./figures/}}

\newcommand{\gettitle}{
Physics of the gluon mass gap}

\newcommand{\getHeidelbergAffiliation}{\affiliation{Institut f\"ur Theoretische Physik, Universit\"at Heidelberg, Philosophenweg 16, 69120 Heidelberg, Germany}}
\newcommand{\getDarmstadtAffiliation}{\affiliation{Institut f\"ur Kernphysik, Technische Universit\"at Darmstadt, Schlossgartenstra{\ss}e 2, 64289 Darmstadt, Germany}}
\newcommand{\getEMMIAffiliation}{\affiliation{ExtreMe Matter Institute EMMI, GSI, Planckstr. 1, 64291 Darmstadt, Germany}}
\newcommand{\getNJUAffiliation}{\affiliation{School of Physics, Nanjing University, Nanjing, Jiangsu 210093, China}}
\newcommand{\getINPAffiliation}{\affiliation{Institute for Nonperturbative Physics, Nanjing University, Nanjing, Jiangsu 210093, China}}

\newcommand{\getVLCAffiliation}
{\affiliation{Department of Theoretical Physics and IFIC, University of Valencia and CSIC, E-46100, Valencia, Spain}}

\begin{document}

\title{\gettitle}

\author{Mauricio N. Ferreira}
\getNJUAffiliation
\getINPAffiliation

\author{Joannis Papavassiliou}
\getVLCAffiliation
\getEMMIAffiliation

\author{Jan M. Pawlowski}
\getHeidelbergAffiliation
\getEMMIAffiliation
  
\author{Nicolas Wink}
\getDarmstadtAffiliation

\begin{abstract}

It has long been known that the gluon propagator in Landau-gauge QCD exhibits a mass gap;
and its emergence has been ascribed to the action of the Schwinger mechanism in the gauge sector of QCD. In the present work, we relate this property to the physical mass gap of QCD by considering two observables associated with confinement and chiral symmetry breaking, namely the confinement-deconfinement transition temperature and the pion decay constant, respectively. It turns out that the first observable is linearly proportional to the gluon mass gap, a fact that allows us to assign a direct physical meaning to this scale. Moreover, we identify three distinct momentum regimes in the gluon propagator,  
ultraviolet, intermediate, and deep infrared, 
and assess their impact on the aforementioned
observables. Both observables are sensitive to the first two regions of momenta, 
where functional approaches essentially coincide, but are insensitive to the third, deep infrared, regime. 
The combined information is used for a simple fit for the gluon propagator, 
all of whose parameters admit a clear physical interpretation. Finally, we discuss how this fit can help us access the intertwined dynamics of confinement and chiral symmetry breaking in QCD-type theories.  

\end{abstract}

\maketitle

\section{Introduction}
\label{sec:Introduction}

In the past two decades, functional methods,
such as Dyson-Schwinger equations (DSEs)~\cite{Alkofer:2000wg, Fischer:2006ub, Binosi:2009qm, Maas:2011se, Cloet:2013jya, Fischer:2018sdj, Huber:2018ned, Ferreira:2023fva}  
and the functional renormalisation group 
(fRG)~\cite{Pawlowski:2005xe, Gies:2006wv, Braun:2011pp, Dupuis:2020fhh, Fu:2022gou},
have evolved into a versatile and quantitative first-principles approach to QCD. 
Within this general framework, the correlation functions of fundamental fields are obtained by solving coupled dynamical equations. 
These relations are one-loop exact within the fRG approach, and two-loop exact within the 
DSE approach. 
In both cases, a gauge-fixed 
formulation of the theory is required, and the Landau gauge constitutes the standard choice due to its many practical and conceptual advantages. 
The physical applications accessible by functional approaches include both static and time-like phenomena, and range from the resonance spectrum and scattering events in low-energy QCD to the phase structure and observables at finite temperature and density. 

Since low-energy QCD is governed by confinement and chiral symmetry breaking, a comprehensive understanding of these emergent phenomena in terms of \textit{gauge-fixed} correlation functions is essential for the understanding of infrared QCD as well as the error control of the aforementioned applications. 

In the present work we contribute to this ongoing task by scrutinising certain key features of the Landau gauge gluon propagator and the associated dressing function. 
In particular, we focus on the presence of a gluon mass gap, $m_\textrm{gap}$, 
and explore its implications on two fundamental QCD quantities, namely the confinement-deconfinement temperature, $T_c$, and the pion decay constant, $f_\pi$. We will show that both $T_c$ and $f_\pi$
are insensitive to variations of the gluon propagator below a physical infrared scale, which signals the effective decoupling of the gluon dynamics below the gluon mass gap.   

Our analysis is guided by the presence of three distinct momentum regimes in the gluon propagator: (${\it uv}$) the ultraviolet regime, where the physics is well-described by (resummed) perturbation theory;  (${\it sc}$) the strongly correlated intermediate momentum regime, which exhibits the dynamical emergence of $m_\textrm{gap}$; 
 and (${\it ir}$) the deep infrared regime,
 where the effective decoupling of the gluonic dynamics occurs. Note that it is this decoupling which signals the presence of a gluon mass gap $m_\textrm{gap}$ in the first place. 
In fact, the transition between these 
momentum regimes 
may be easily visualised, 
thanks to the topological 
features exhibited by the gluon dressing function,
namely 
a maximum 
and two inflection points. 

We argue that state-of-the-art functional approaches provide reliable quantitative results for regimes (${\it uv}$) and (${\it sc}$), while regime (${\it ir}$) leaves no imprint on the observables considered here. The positions of these special points define three momentum scales, all of which are proportional to the mass gap. In what follows we will interpret these  scales as ``proxies'' that mark the transition between regions of momenta dominated by distinct physical effects,
 namely asymptotic freedom, 
 Schwinger mechanism, and 
 infrared decoupling. 

 The article is organised as follows. 
In \Cref{sec:Schwinger+massgap} we discuss the dynamics of the Schwinger mechanism in the gauge sector, the consequences of the dynamical emergence of the gluon mass gap $m_\textrm{gap}$, and the three momentum regimes \textit{(uv,sc,ir)} introduced above. In \Cref{sec:GluonGapObservables} we discuss the non-perturbative computation of $T_c$ and $f_\pi$ and their (in-)dependence on the regimes \textit{(uv,sc,ir)}. In \Cref{sec:GluonMassGap} we use our findings for a minimal fit for the gluon propagator, where all fitting parameters directly carry key properties of the three regimes. This also allows us to establish a linear relation between $T_c$ and the gluon mass gap. We also discuss how this fit can be put to work for assessing the intertwined dynamics between confinement and chiral symmetry breaking in QCD-type theories. Our results are briefly reviewed in \Cref{sec:Conclusion}. 
Finally, certain technical points are discussed in four Appendices.

\section{Schwinger mechanism 
and the gluon mass gap}
\label{sec:Schwinger+massgap} 

It is well-established by now that a mass gap in QCD is dynamically generated within covariant gauges via the Schwinger mechanism~\cite{Schwinger:1962tn, Schwinger:1962tp, Jackiw:1973tr, Eichten:1974et, Smit:1974je, Cornwall:1981zr}; for a recent review on the subject,  see~\cite{Ferreira:2025anh}. 
This mechanism is operative 
both in the Landau gauge
($\xi=0$), the case most widely 
studied in the literature, as well as finite gauge-fixing parameter $\xi\neq 0$
~\cite{Aguilar:2015nqa,Aguilar:2016ock,Napetschnig:2021ria}. In addition, it persists for QCD with a sufficiently low number of dynamical quarks, including the case when the current quark masses acquire their physical values~\cite{Aguilar:2013hoa}. 

Technically, the occurrence of the Schwinger mechanism requires a special type of irregularities in the fully-dressed three-gluon and the ghost-gluon vertices; 
specifically, at low momentum transfer, $p^2 \to 0$, these vertices display {\it longitudinally-coupled} massless poles that carry color. These features are purely non-perturbative:
they arise when two gluons or a ghost-antighost pair merge to form a {\it colored} composite excitation of vanishing mass~\cite{Ferreira:2025anh}. 
The effects of these irregularities are crucial in the infrared, but weaken rapidly in the ultraviolet, leaving no trace at the perturbative level. 
 
We denominate \textit{gluon mass gap} the renormalisation group invariant (RGI) mass scale emerging from the action of the Schwinger mechanism, and denote it by $m_{{\rm gap}}$. 
The situation is physically akin to the emerging RGI mass scale $m_\chi$ associated with dynamical chiral symmetry breaking (DCSB), see the reviews~\cite{Alkofer:2000wg, Fischer:2006ub, Binosi:2009qm, Maas:2011se, Cloet:2013jya, Fischer:2018sdj, Huber:2018ned, Ferreira:2023fva, Pawlowski:2005xe, Gies:2006wv, Braun:2011pp, Dupuis:2020fhh, Fu:2022gou}. 
Indeed, as in the case of the gluon mass gap, the effects of DCSB are undetectable at any finite order in perturbation theory. Moreover, if integrating out momentum modes successively as in the fRG, no signature of spontaneous chiral symmetry breaking is present above the onset cutoff scale $k_\chi\approx 500$\,GeV of DCSB, see e.g.~\cite{Gies:2002hq, Mitter:2014wpa, Braun:2014ata, Cyrol:2017ewj, Goertz:2024dnz, Fu:2025hcm}. Finally, while DCSB leaves its trace on many observables, such as the chiral condensate, meson masses, decay constants, and the chiral transition temperature, no unique value can be directly ascribed to $m_\chi$; 
in fact, chiral scales may be defined via any of these observables.    

Similarly, while no direct value may be assigned to it, $m_{{\rm gap}}$ leaves its trace on observables, most notably in the glueball masses and the confinement-deconfinement temperature of both Yang-Mills theory and QCD. 
This suggests a mass scale of about 1\,GeV, and numerous studies indicate that the dynamics of the Schwinger mechanism occurs indeed at this scale, see~\cite{Ferreira:2025anh} 
and references therein. 
Finally, in analogy to the appearance of the constituent quark mass in the quark propagator $\langle q(p) \bar q (-p)\rangle_c$, the gluon mass gap $m_{{\rm gap}}$ leads to a gapped gluon propagator, 
$\langle A^a_\mu(p) \, A^b_\nu(-p)\rangle_c$; here the subscript ``$c$'' indicates the connected part of the two-point function.

In the Landau gauge, the gluon propagator is completely transverse, 
\begin{subequations}
\label{eq:GluonProp}
\begin{align} 
 \langle A^a_\mu(p) \, A^b_\nu(-p)\rangle = G_A(p) \,\delta^{ab}\,\Pi^\bot_{\mu\nu}(p) \,,
    \label{eq:GluonPropExplicit}
\end{align}
where 
\begin{align}
    \Pi^\bot_{\mu\nu}(p) =\delta_{\mu\nu}- \frac{p_\mu p_\nu}{p^2}\,,
\end{align}
is the standard transverse projection operator. Note that in~\cite{Ferreira:2025anh} and further DSE works,  
the scalar part 
$G_A(p)$ is denoted by $\Delta(p)$. The scalar propagator  
$G_A(p)$ can be parametrised as a product of its classical counterpart, $1/p^2$, and the propagator dressing $1/Z_A(p)$, 
\begin{align}
G_A(p)=   \frac{1}{Z_A(p)} \frac{1}{p^2} \,.
\label{eq:ScalarGA}
\end{align}
\end{subequations} 
This notation is commonly used in functional renormalisation group studies, see e.g.,~\cite{Cyrol:2016tym}. 

\begin{figure}[t]
  \includegraphics[width=0.45\textwidth]{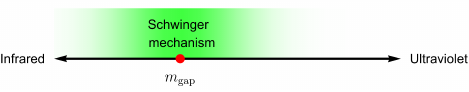}  
  \caption{The three momentum regimes in Landau-gauge QCD:  The gluon mass gap, $m_{{\rm gap}}$, is generated by the 
  dynamics associated with the Schwinger mechanism, taking place 
  in the green shaded region of momenta, and its 
  location is indicated by the red dot. 
  On the right of this point, 
  asymptotic freedom eventually sets in, 
  and  perturbation theory provides  
  an accurate description. Left of 
  $m_{{\rm gap}}$ is the infrared regime, 
  where the gluonic dynamics 
  gradually attenuate and finally 
  decouple. In \Cref{fig:gluon_dressing_scales}, we show the gluon dressings in Yang-Mills theory and 2+1 flavour QCD, using the same color coding.\hspace*{\fill}}
  \label{fig:cartoon_SM}
\end{figure}

In the deep ultraviolet, the theory is perturbative, and the gluon propagator displays the typical logarithmic scaling,
leading to the decay of $Z^{-1}_A(p)$. The activation of the Schwinger mechanism takes place in the vicinity of 1\,GeV, see \Cref{fig:cartoon_SM}, below which the gluon propagator decouples gradually from the dynamics. 
In this regime, the form of $G_A(p)$, as well as of other ghost-gluon correlation functions, is governed by the dynamics of the infrared enhanced ghost. 
Note, however, that the matter sector of QCD is not affected directly by the infrared ghost dynamics, but only by the gradually decoupling gluon.  

%
\begin{figure*}[t]
  \begin{subfigure}[t]{.45\linewidth}
\includegraphics[width=\linewidth]{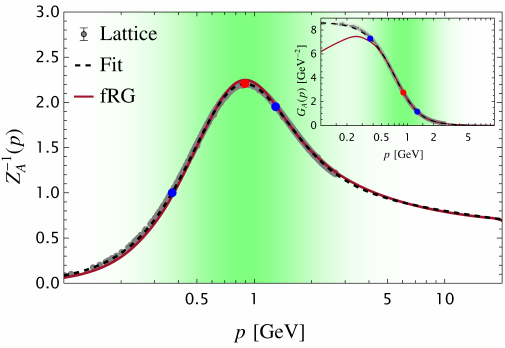}
		\caption{Gluon dressing function n Yang-Mills theory: lattice data \cite{Aguilar:2021uwa} (data points), fRG data \cite{Cyrol:2016tym} (red line), fit (dashed black line) obtained from both data sets, see \labelcref{eq:GlobalFitGluon} together with \Cref{tab:fit_params} in \Cref{sec:ScreeningMassFits}.\hspace*{\fill}}
		\label{fig:YMdressing}
	\end{subfigure}%
	\hspace{0.05\linewidth}%
     \begin{subfigure}[t]{.45\linewidth}
\includegraphics[width=\linewidth]{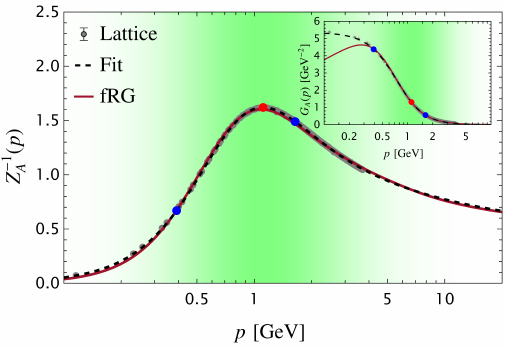}
		\caption{Gluon dressing in 2+1 flavour QCD: lattice data \cite{Aguilar:2019uob} (data points), fRG data \cite{Fu:2025hcm} (red line), fit (dashed black line) obtained from both data sets, see \labelcref{eq:GlobalFitGluon} together with \Cref{tab:fit_params} in \Cref{sec:ScreeningMassFits}. \hspace*{\fill}}
		\label{fig:2+1QCDdressing}
	\end{subfigure}%
  \caption{Dressing $p^2\,G_A(p)$ 
  of the gluon propagator $G_A(p)$  \labelcref{eq:GluonProp} as a function of momenta. We show the dressing in Yang-Mills theory, \Cref{fig:YMdressing}, and in 2+1 flavour QCD, \Cref{fig:2+1QCDdressing}. The insets show the respective propagators. We use lattice data, functional QCD data, and a fit (dashed black line) obtained from both data sets, see \labelcref{eq:GlobalFitGluon} with \Cref{tab:fit_params} in \Cref{sec:ScreeningMassFits}. The green-shaded area indicates that in \Cref{fig:cartoon_SM}: we have used the inflection points \labelcref{eq:InflectionUV,eq:InflectionIR}, see \Cref{tab:Inflection}, as proxies for the ultraviolet and infrared boundaries of the dynamics regime of the Schwinger mechanism, and the peak position \labelcref{eq:peak} as a proxy of the gluon mass gap. The similarity between Yang-Mills theory and 2+1 flavour QCD indicates the gluonic nature of the green-shaded areas, dominated by the dynamics of the Schwinger mechanism.\hspace*{\fill}}
  \label{fig:gluon_dressing_scales}
\end{figure*}
%
As decoupling typically takes between half to one order of magnitude to occur, the respective decoupling scale is expected to take a value between 0.5 - 0.1\,GeV. Moreover, its value will depend on the specific definition.  
This qualitative picture is supported by the computational results in pure Yang-Mills theory, see e.g.,~\cite{Cyrol:2016tym, Pawlowski:2022oyq} for comprehensive functional renormalisation group studies, and \cite{Huber:2020keu} for a respective DSE study, for a rather complete survey of functional as well as lattice data see the recent reviews~\cite{Huber:2018ned, Dupuis:2020fhh, Ferreira:2023fva, Ferreira:2025anh}. 

In \Cref{fig:gluon_dressing_scales} we show representative fRG data~\cite{Cyrol:2016tym} (red continuous line) and lattice data~\cite{Aguilar:2021okw} (data points) for the gluon dressing; 
the corresponding gluon propagators are shown in the inlay. 
The propagator dressings are normalised to unity at the renormalisation scale $\mu = 4.3$\,GeV used in the computations of the present work, except in \Cref{sec:TfpimGap}, where $\mu$ will be varied. We have  
\begin{align} 
Z_A(\mu)=1\,,\qquad Z_q(\mu)=1\,,\qquad \mu=4.3\,\textrm{GeV}\,, 
\label{eq:RG-Conditions}
\end{align}
where we also have added the renormalisation condition for the quark propagator dressing $Z_q(p)$, see \labelcref{eq:QuarkProp}.   

In \Cref{fig:gluon_dressing_scales}, we also included a fit to the combined lattice and functional data (dashed black line); the specifics of the fit will be discussed in \Cref{sec:ScreeningMassFits}. 
The quantitative agreement in the ultraviolet and green regimes is indicative of the maturity that functional approaches have reached in the past two decades. 
The gluon propagators from state-of-the-art functional approaches (both DSE and fRG) agree quantitatively with each other and with the reference lattice data above $~500$\,MeV; see \Cref{fig:gluon_dressing_scales} for specific solutions in Yang-Mills theory and $2+1$ flavour QCD. In the deep infrared, functional results in the literature depend on the details of the ghost dynamics, while lattice results show a saturation of the gluon propagator: the gluon propagator reaches a 
nonvanishing value at $p=0$,  $G_A(0) =1/m_\textrm{sat}^{2}$,
(decoupling solutions/lattice). Functional results in the deep infrared vary between lattice-type solutions and that with a vanishing gluon propagator or diverging $m_\textrm{sat}$, $G_A(0) =0$ (scaling solutions). By its definition, the value of $m_\textrm{sat}$ depends on the renormalisation point $\mu$. Instead, the gluon mass gap is directly linked to observables, and will be defined as an RGI quantity, in analogy to the chiral mass scale $m_\chi$.  

With this general qualitative picture in mind, we now take a closer look at characteristic signatures for the Schwinger mechanism, leading to the three momentum regimes sketched in \Cref{fig:gluon_dressing_scales}. 
Qualitatively, these three regimes can be extracted from the gluon dressing function depicted in \Cref{fig:gluon_dressing_scales}. 
In the following, we discuss emergent features of the Schwinger mechanism, starting in the perturbative ultraviolet, and finally approaching the non-perturbative infrared regime:\\[-2ex]

({\it i}) The first clear signature is a qualitative change in the momentum dependence of the propagator dressing 
$1/Z_A(p)$: while it is a 
convex function for perturbative momenta, it changes its curvature and becomes concave. 
This change is signalled by the corresponding ultraviolet inflection point of $1/Z_A(p)$ with the location $p_{\textrm{in}}^+$,  
\begin{align} 
\left. \frac{\partial^2 Z^{-1}_A(p)}{\partial p^2} \right|_{p=p_{\textrm{in}}^+} =0\,.
\label{eq:InflectionUV}
\end{align}
For the lattice and fRG data in \Cref{fig:gluon_dressing_scales}, this inflection point is located at $p_{\textrm{in}}^+ \approx 1.3$~GeV for Yang-Mills theory and $1.6$~GeV for $2+1$ flavor QCD, and 
is indicated by a blue dot.\\[-2ex] 

({\it ii}) The second signature of the gluon mass gap is the peak of the dressing function $1/Z_A$. Its location $p_\textrm{peak}$ is defined as   
\begin{align} 
p_\textrm{peak}:\quad \max_p\left[{Z^{-1}_A(p)}\right] = 
Z^{-1}_A (p_\textrm{peak}) \,, 
\label{eq:peak}
\end{align}
which translates readily into 
\begin{align} 
\left. \frac{\partial Z_A(p)}{\partial p} \right|_{p=p_{\textrm{peak}}} =0\,.
\label{eq:PeakPosition}
\end{align}
Note that such a peak is not present in the dressing of a classical massive propagator $1/(p^2+m^2)$. Its presence in the gluon originates from the interplay between the logarithmic decay of $1/Z_A(p)$ in the ultraviolet and the suppression caused by the generation of $m_{{\rm gap}}$ in the infrared. For the lattice and fRG data depicted in \Cref{fig:gluon_dressing_scales}, the peak position is at $p_{\textrm{peak}} \approx 900$~MeV and $1.1$~GeV for $N_f = 0$ and $N_f = 2+1$, respectively, and is depicted as a red dot.\\[-2ex] 

({\it iii}) Finally, 
there is a third notable signature, namely  
a second inflection point, 
whose location 
$p_{\textrm{in}}^-$ is 
to the left 
of the peak position $p_{\textrm{peak}}$. We denote this as the ``infrared inflection point''. 
In complete analogy to \labelcref{eq:InflectionUV} we define 
\begin{align} 
\left. \frac{\partial^2 Z^{-1}_A(p)}{\partial p^2} \right|_{p=p_{\textrm{in}}^-} =0\,.
\label{eq:InflectionIR}
\end{align}
For the lattice and fRG data in \Cref{fig:gluon_dressing_scales}, this inflection point is located at $p_{\textrm{in}}^- \approx 370$~MeV and $390$~MeV, for Yang-Mills and $2+1$ flavour QCD, respectively, and as its ultraviolet counterpart $p_{\textrm{in}}^+$, is indicated by a blue dot. \\[-2ex] 

This concludes our discussion of the signatures of the Schwinger mechanism in the gluon propagator. We have argued that all three properties are clearly linked to $({\it i})$ the onset of the Schwinger dynamics roughly below $p_{\textrm{in}}^+$, $({\it ii})$ the mass gap itself, and
$({\it iii})$ the decoupling of the gluon dynamics with $p_{\textrm{in}}^-\approx p_\textrm{dec}$. 

%
\begin{figure*}[t]
  \begin{subfigure}[t]{.48\linewidth}
\includegraphics[width=\linewidth]{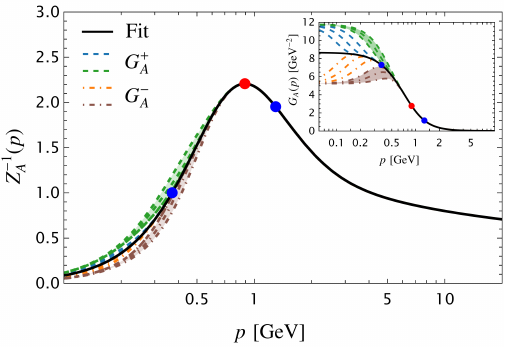}
		\caption{Yang-Mills theory  }
		\label{fig:IR-DeformationsYM}
	\end{subfigure}%
	\hspace{0.5cm}%
     \begin{subfigure}[t]{.48\linewidth}
\includegraphics[width=\linewidth]{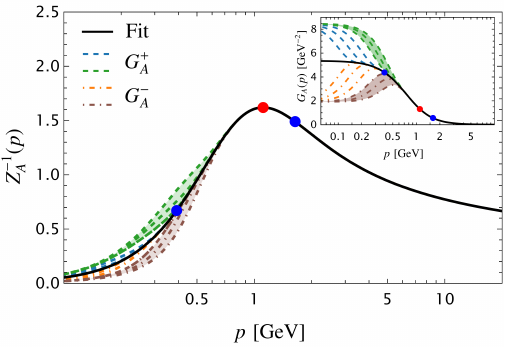}
		\caption{2+1 flavour QCD}
		\label{fig:IR-DeformationsQCD}
	\end{subfigure}%
  \caption{Gluon dressings and propagators (inlays) in Yang-Mills theory, \Cref{fig:IR-DeformationsYM}, and 2+1 flavour QCD, \Cref{fig:IR-DeformationsQCD}: Fits  \labelcref{eq:GlobalFitGluon} with the parameters in \Cref{tab:fit_params} (continuous black line), infrared deformations \labelcref{eq:GA_vars} in \Cref{app:prop_vars}. The green- and brown-shaded areas indicate the deformations that have an impact on the observables in \Cref{fig:Observables-DeformationsYM+QCD}.\hspace*{\fill}}
  \label{fig:IR-DeformationsYM+QCD}
\end{figure*}
%

%
\begin{figure*}[t]
  \begin{subfigure}[t]{.48\linewidth}
\includegraphics[width=\linewidth]{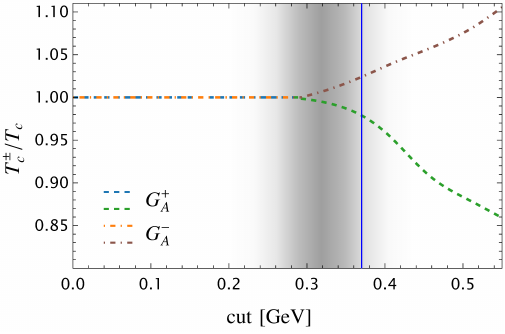}
		\caption{Critical temperature $T^{\pm}_c$ in Yang-Mills theory, computed with the IR-deformations of the gluon propagator in \Cref{fig:IR-DeformationsYM}, and normalised by the physical $T_c$.  \hspace*{\fill}}
		\label{fig:Tc-Deformations}
	\end{subfigure}%
	\hspace{0.5cm}%
     \begin{subfigure}[t]{.48\linewidth}
\includegraphics[width=\linewidth]{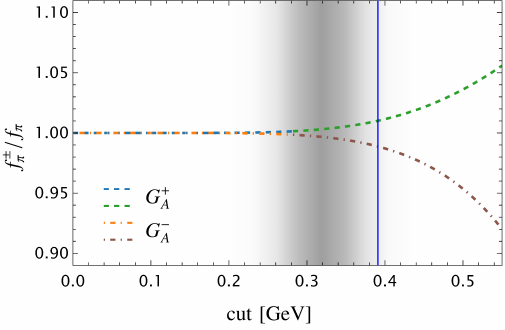}
	\caption{Pion decay constant $f_\pi^\pm$ in 2+1 flavour QCD, computed with the IR-deformations of the gluon propagator in \Cref{fig:IR-DeformationsQCD}, and normalised by the physical $f_\pi$.  \hspace*{\fill}}
    \label{fig:fpi-Deformations}
    \end{subfigure}
  \caption{Confinement-deconfinement temperature $T^\pm_\textrm{conf}$ in Yang-Mills theory, and pion decay constant in 2+1 flavour QCD, as functions of the infrared deformation of the corresponding gluon propagators, 
  displayed in \Cref{fig:IR-DeformationsYM} and  \Cref{fig:IR-DeformationsQCD}. Both observables are normalised by their physical values $T_c$ and $f_\pi$, respectively. The grey-shaded area indicates the regime in which the observables are getting affected by the infrared deformations of the gluon propagators in \Cref{fig:IR-DeformationsYM+QCD}. The infrared inflection points \labelcref{eq:InflectionIR} are indicated by vertical blue lines, for their values see \Cref{tab:Inflection}. \hspace*{\fill}}
  \label{fig:Observables-DeformationsYM+QCD}
\end{figure*}
%

The decoupling scale $p_\textrm{dec}$ can be extracted directly from observables as the RGI scale, below which these observables are insensitive to the (change of the) momentum dependence of the gluon. This is the typical decoupling known from ``massive'' theories; in principle,  
we expect $p_\textrm{dec}$ to be proportional to but smaller than the gluon mass gap, 
$p_\textrm{dec} < m_\textrm{gap}$. 

\begin{table}[t]
\vspace{.2cm}
\begingroup
\renewcommand{\arraystretch}{1.3}
\setlength{\tabcolsep}{6pt} %
\begin{tabular}{|c||c|c|c|}
	\hline
	$N_f$  & $p_{\textrm{in}}^+[\textrm{MeV}]$ & $p_{\textrm{peak}}[\textrm{MeV}]$ & 
     $p_{\textrm{in}}^-[\textrm{MeV}]$ \\
	\hline
	$0$  & 
    1294 & 
    892 & 
    370  \\
    \hline 
    $2+1$  & 
    1638 & 
    1112 & 
    391 \\
	  \hline
\end{tabular}
\caption{Values of the UV inflection point $p_{\textrm{in}}^+$ in \labelcref{eq:InflectionUV}, the peak position $p_{\textrm{peak}}$ in \labelcref{eq:PeakPosition}, and the IR inflection point $p_{\textrm{in}}^-$ in \labelcref{eq:InflectionIR}. We list them for the Yang-Mills theory ($N_f = 0$), and $2+1$ flavour QCD. \hspace*{\fill}}
\label{tab:Inflection}
\endgroup
\end{table}

In the present work we use two observables related to (${\it a}$) confinement and 
$({\it b})$ DCSB, for providing estimates for the decoupling scale. More specifically we use 
\begin{itemize} 
\item[(${\it a}$)] Confinement-deconfinement phase transition temperature $T_c$,  
\item[(${\it b}$)] Pion decay constant $f_\pi$. 
\end{itemize} 
In summary, the infrared dynamics driven by the Schwinger mechanism leads to the natural division of the gluon propagator momentum range into three regimes: 
the first is the perturbative ultraviolet regime, $p^2 \gtrsim (p^+_\textrm{in})^2$, which is roughly bounded by the ultraviolet inflection point. 
The second is the regime that carries the dynamics of the Schwinger mechanism, with $p^2_\textrm{dec}\lesssim p^2\lesssim (p^+_\textrm{in})^2$. 
Finally, we have the deep infrared regime, $p^2\lesssim p^2_\textrm{dec}$, where the gluon dynamics has decoupled. 
All the above points are depicted in \Cref{fig:cartoon_SM} and \Cref{fig:gluon_dressing_scales}, and their values are collected in \Cref{tab:Inflection}.

\section{Gluon mass gap and observables}
\label{sec:GluonGapObservables} 

The dynamical gluon mass gap leaves direct and indirect traces on a plethora of observables, for early works see in particular~\cite{Parisi:1980jy, Cornwall:1981zr, Bernard:1981pg, Bernard:1982my, Donoghue:1983fy, Mandula:1987rh}, for recent state-of-the-art functional results and further developments see the reviews~\cite{Alkofer:2000wg, Fischer:2006ub, Binosi:2009qm, Maas:2011se, Cloet:2013jya, Fischer:2018sdj, Huber:2018ned, Ferreira:2023fva, Pawlowski:2005xe, Gies:2006wv, Braun:2011pp, Dupuis:2020fhh, Fu:2022gou}.  

As discussed at the end of \Cref{sec:Schwinger+massgap}, 
in the present work we shall consider two of them, namely 
(${\it a}$) $T_c$ and  (${\it b}$) $f_\pi$, 
linked to confinement and 
chiral symmetry, respectively.
We have chosen these two particular 
examples because they are representative of a large 
class of observables, 
whose definition involves 
integrated combinations of gauge-fixed momentum-dependent correlation functions. For the cases in hand, 
this entails that gapped propagators, such as the gluon propagator, 
are suppressed in the infrared, due to the corresponding momentum integration, namely with $d^4 p$ for $f_\pi$, and $d^3 p$ for the transition temperature $T_c$. Moreover, both $T_c$ and $f_\pi$ have relatively simple loop momentum representations. 
In fact, in Yang-Mills theory,  
$T_c$ only depends on the ghost and gluon propagators, while in QCD with dynamical quarks it also involves quark propagators. 

The upshot of our considerations is that both observables are insensitive to deformations of the gluon propagator below the decoupling scale; this happens because, effectively, this class of changes leaves the gluon mass gap {\it invariant}.
Instead, modifications above the decoupling scale amount to effectively changing the gluon mass gap; as a result, the two observables, $T_c$ and $f_\pi$, change correspondingly.

The gluon propagators employed are shown in \Cref{fig:IR-DeformationsYM+QCD}, for both Yang-Mills theory and 2+1 flavour QCD. 
The deformations shown are defined in \labelcref{eq:GA_vars} of \Cref{app:prop_vars}, and are based on the combined fits of lattice and functional data \labelcref{eq:GlobalFitGluon}, with the relevant parameters in \Cref{tab:fit_params}. 
For the guidance of the reader, and for a better appreciation of the respective computations, we display in advance the results in \Cref{fig:Observables-DeformationsYM+QCD}: 
the complete stability of the observables under changes of the gluon propagator for momenta $p_\textrm{dec} \approx p_{\textrm{in}}^-$ is manifest. 
Instead, when modifications above the decoupling scales are effectuated, the observables undergo visible changes. 

In \Cref{sec:FullPolyakov,sec:ChiralGap} we provide the details of the respective analyses: 
In \Cref{sec:FullPolyakov} we study $T_c$ within the DSE approach, where it can be computed directly from the gluon and ghost propagators. 
In \Cref{sec:ChiralGap} we consider the gap equation, and the impact of the gluon mass gap on its solution, as well as on attendant observables, such as the pion decay constant.

\subsection{Confinement-deconfinement phase transition and the gluon mass gap}
\label{sec:FullPolyakov} 

\begin{figure*}[t]
	\centering
	\includegraphics[width=.7 \textwidth]{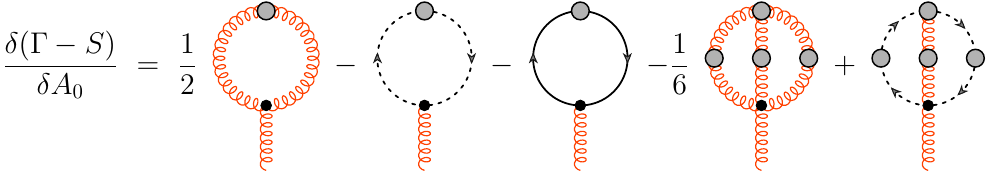} 
	\caption{Functional background field DSE. Full propagators are indicated by lines with grey circles, while  classical vertices by small black circles. In contradistinction to its quantum-field counterpart, this DSE also features a two-loop ghost-gluon term, with the four-point vertex $S^{(4)}_{ac\bar c \bar A_0}$. \hspace*{\fill}}
	\label{fig:DSE-A}
\end{figure*}
The functional approach to the physics of the confinement-deconfinement phase transition has been put forward in \cite{Braun:2007bx,Fister:2013bh}. In functional approaches, the rôle of the standard order parameter is assumed by the expectation value of the traced Polyakov loop, namely the Wilson line in the temporal direction, 
\begin{align}
 L = \frac{1}{N_c} \textrm{Tr}_f 
 P(\vec x)\,,\qquad P(\vec x) = {\cal P} e^{{i \,g_s \int\limits_0^{\beta}\!\!d \tau \,
 A_0(\tau,\vec x) }}\,. 
 \label{eq:TracedP}
 \end{align} 
In \cite{Braun:2007bx, Fister:2013bh} this observable has been related to the gauge-invariant expectation value of the background temporal gauge field, $\langle A_0\rangle$. For that purpose, we parametrise 
 \begin{align}
 P(\vec x) = e^{2 \pi i \phi(\vec x)}\,. 
\label{eq:PolloopMain}
\end{align}
The algebra field $\phi$ carries the physics information of the Polyakov loop: the related eigenmode field $\varphi$ of $\phi$ is gauge invariant, and defines an order parameter, see \labelcref{eq:EVPolloop}. 
The field $\varphi$ is directly related to the eigenmode field of the temporal background gauge field, $\bar A_0$, see \labelcref{eq:EVvarphi}.
By virtue of these relations,
the solution of the background equation for $\bar A_0$, to be denoted by $\langle \bar A_0\rangle$, is elevated to an order parameter itself;  
for details, see \Cref{app:FunConf-Deconf}. 
Due to its close relation to the standard order parameter $\langle L\rangle$, in most cases  $L[\langle \bar A_0\rangle]$ is used instead of $\langle \bar A_0\rangle$. 
These two different order parameters have been related in \cite{Herbst:2015ona}, and the functional results for $\langle L[A_0]\rangle$, derived from $L[\langle A_0]\rangle$, are in quantitative agreement with lattice results. 

In summary, the physics of the confinement-deconfinement phase transition can be studied directly within functional QCD: 
the computation of the order parameter potential for $\bar A_0$ gives us access to the critical temperature via $\langle \bar A_0\rangle$. 
Note that this particular background enters directly in various additional observables, sensitive to the confinement-deconfinement physics, such as the fluctuations of conserved baryon charges; 
for results for SU($N_c$) see \cite{Braun:2010cy}, for results in full QCD see e.g.,~\cite{Braun:2009gm, Fischer:2013eca, Fischer:2014ata, Fu:2019hdw, Lu:2023mkn, Lu:2025cls}. 
Given the importance of $\bar A_0$
to the present work, we briefly review its main properties in \Cref{app:FunConf-Deconf} and also refer to~\cite{Lu:2025cls} for a recent detailed account in 2+1 flavour QCD.

\subsubsection{Polyakov-loop effective potential in functional approaches}
\label{sec:Veff}

It is left to compute the effective potential $V_\textrm{eff}$ of the temporal background field $\varphi(A_0)$,
whose minima define the order parameter. For constant gauge fields $A_0$, we find 
\begin{align} 
\frac{\partial  V_\textrm{eff}(\varphi)} {\partial \varphi} = 0\,,\qquad V_{\textrm{eff}}(\varphi)  = \frac{T}{{\cal V}_3}\Gamma[A_0]\,, 
\label{eq:EoMvarphi}
\end{align}
and $\varphi(A_0)$ is given by \labelcref{eq:EVvarphi}. Here, 
${\cal V}_3$ denotes the spatial  
volume. \Cref{eq:EoMvarphi} can be computed from the $A_0$-DSE for the background field effective action, see \cite{Fister:2013bh}. It is depicted in \Cref{fig:DSE-A} for full QCD; pure Yang-Mills theory is obtained by omitting the quark loop. From now on we will drop the ``bar'', and refer to the background field simply as $A_0$, since we evaluate the effective action for vanishing fluctuating field, $a_\mu=0$. Some basic properties of the background field approach are provided in \Cref{app:FunConf-Deconf}; for more details we refer to the literature, and in particular to the recent comprehensive account for 2+1 flavour QCD at finite temperature and density given in \cite{Lu:2025cls}. 

The crucial input in the $A_0$-DSE in \Cref{fig:DSE-A} are the ghost, gluon, and quark propagators, and the approach allows us to unravel the direct relation between the critical temperature $T_c$ and the gluon mass gap. For the structural and illustrational purposes of the present work we will drop the two-loop terms in \Cref{fig:DSE-A}. Note that the quantitative accuracy of this approximation has been tested both in \cite{Fister:2013bh} and in the many applications since then. 

A further simplification is the use of vacuum propagators instead of thermal ones. While 
for temperatures $T\lesssim T_c$
this is quantitatively accurate for the ghost, the gluon mass gap receives thermal corrections, see \cite{Cyrol:2017qkl}, which have a subleading impact for $T\lesssim T_c$. Finally, we shall concentrate on SU(3) Yang-Mills theory, which exhibits a first order phase transition, with a uniquely defined critical temperature; instead, in full QCD we have a crossover, due to the explicit center symmetry breaking induced by the quarks. 

The only inputs required for 
the computation 
of the first two loop diagrams in 
\Cref{fig:DSE-A} 
are the ghost and gluon propagators. The latter one has been parametrised in \labelcref{eq:GluonProp}, and its dressing $1/Z_A(p)$ is depicted in \Cref{fig:gluon_dressing_scales}. The ghost propagator is parametrised as 
\begin{subequations} 
\label{eq:GhostProp}
\begin{align}
\langle c(p) \bar c(q) \rangle_c = G_c(p)\, \delta^{ab} \,(2 \pi)^4 \delta(p+q)\,,
\label{eq:Cprop} 
\end{align}
with the scalar part $G_c(p)$. Similarly to the gluon propagator, it can be split into the classical part $1/p^2$ and the propagator dressing, 
\begin{align} 
G_c(p) = \frac{1}{Z_c(p)}\frac{1}{p^2}\,.
\label{eq:Zc} 
\end{align}
\end{subequations} 
Functional results and lattice data for the ghost dressing $1/Z_c(p) $ are depicted in \Cref{fig:ghost_dressing} in \Cref{app:FitGhost}, where we also provide an analytic fit.  

Denoting by $V_A$ and $V_c$
the contribution of the gluon and ghost loop, respectively, 
we have
\begin{align} 
V_\textrm{eff}(\varphi)= V_A(\varphi)+ V_c(\varphi)\,. 
\end{align} 
The respective color traces in 
$V_A$ and $V_c$
can be performed in terms of the eigenvectors $\hat e_j$ and the respective eigenmodes $\nu_j$ defined in \labelcref{eq:EVPolloop}. There are no off-diagonal terms, and the potentials $V_{i}$ with $i=A,c$ can be represented in terms of the mode potential, $V_{i,\textrm{mode}}$, with 
\begin{align}\nonumber 
V_{i,\textrm{mode}}(\nu_j) \propto &\, T\sum_{n\in \mathbbm{Z}} \int \frac{d^3 p}{(2 \pi)^3} \, 
    2 \pi T(n+\nu_j)\\[1ex] 
    &\, \hspace{1cm}\times G_i\bigl(4 \pi^2 T^2(n+\nu_j)^2+\boldsymbol{p}^2\bigr)\,. 
\label{eq:ModePotenial}
\end{align}
The full potential is then given by a sum over all eigenvalues and both fields. In SU(3) this leads us to 
\begin{align} 
V_\textrm{eff}(\varphi_3,\varphi_8)= \sum_{i,n} c_i\, V_\textrm{i,mode}(\nu_n)\,, 
\label{eq:VeffVmode}
\end{align}
where $\varphi_3$ and 
$\varphi_8$ are the field components in the Cartan subalgebra of SU(3) and the $\nu_j$ are linear combinations of $\varphi_3,\varphi_8$, see \labelcref{eq:nu-varphi}.
The coefficients $c_i$ account for the multiplicity of the fields $i=A,c$ in the DSE: for the gluon and ghost parts they may be read off from \Cref{fig:DSE-A}, and by counting the degrees of freedom. For the gluon we have $1/2 * 4=2$ (four polarisations) and for the ghost we have $-1$ from the loop, so that 
\begin{align} 
c_A=2\,,\qquad c_c=-1\,.
\label{eq:ci}
\end{align}
\Cref{eq:VeffVmode} allows for rather direct access to the relation between the critical temperature and the gluon mass gap. All mode potentials are the same for a given field $i=A,c$, and the computation reduces to that of two mode potentials $V_{i,\textrm{mode}}(\varphi)$, and then summing over them in \labelcref{eq:VeffVmode}.

\subsubsection{Confinement and the gluon mass gap}
\label{sec:VeffMassgap}

The importance of the gluon mass gap can already by assessed at the level of a perturbative one-loop computation. 
To that end, we evaluate the first two loops in \Cref{fig:DSE-A} using classical (tree-level) propagators, whose common scalar part, $G^{(0)}_i$, is given by 
\begin{align}
G_i^{(0)} =\frac{1}{4 \pi^2 T^2(n+\nu_j)^2+\boldsymbol{p}^2}\,, 
\qquad i=A,c \,.
\end{align}
Since the ghost and gluon propagators are equal at this level of approximation, the perturbative mode potentials for ghost and gluon are equal, $V_{i,\textrm{mode}}=V_\textrm{mode}$.

\begin{figure}[t]
  \includegraphics[width=0.45\textwidth]{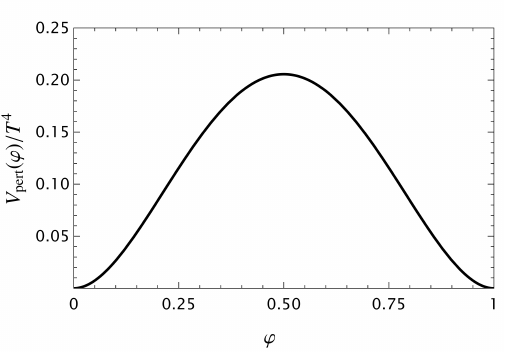}
  \caption{Perturbative potential $V_{\textrm{pert}}(\varphi)$ in SU(2) Yang-Mills theory, see \labelcref{eq:VeffPert}.\hspace*{\fill}}
  \label{fig:Vpert}
\end{figure}

Considering only the non-vanishing eigenvalues \mbox{$\nu_\pm = \pm \varphi_3$}, we get 
\begin{align}
\hspace{-.1cm}V_i(\varphi)=c_i\bigl[ V_\textrm{mode}(\varphi)+V_\textrm{mode}(-\varphi)\bigr]=2c_i V_\textrm{mode}(\varphi).
\label{eq:Vmode}
\end{align} 
Adding up both potentials leads us to  
\begin{align}
 V_{\textrm{pert}}(\varphi)= 2 \,V_A(\varphi) +V_c(\varphi)=2\, V_\textrm{mode}(\varphi)\,, 
\label{eq:VAVcPert}
\end{align}
with $V_A(\varphi) =2\, V_\textrm{pert}(\varphi)$ and $V_c(\varphi) =- V_\textrm{pert}(\varphi)$ \cite{Gross:1980br, Weiss:1980rj}. 
For SU(2) we are led to 
\begin{align}
  V_{\textrm{pert}}(\varphi)=
 \frac{\pi^2}{48} T^4 \left[ 4 \left( \tilde \varphi -\frac12\right)^2 -1\right]^2\,,
\label{eq:VeffPert}
\end{align}
with $\varphi=\varphi_3$ and 
\begin{align}
\tilde \varphi = \varphi\,\textrm{mod}\,1\,. 
\end{align}
The potential $V_{\textrm{pert}}$ in  \labelcref{eq:VeffPert} is displayed in \Cref{fig:Vpert}. 
The minimum of the perturbative potential is at vanishing gauge field, $A_0=0$, that is $\tilde \varphi=0$. 
This leads to a traced Polyakov loop $L[\varphi=0]=1$ for all gauge groups. 
A simple case is SU(2) Yang-Mills theory, where the traced SU(2) Polyakov loop $L[\varphi]= \cos (\pi \varphi)$. 
Evidently, $L=1$ signals deconfinement, see \labelcref{eq:exptrL}. 
Finally, we remark that \labelcref{eq:VeffPert} already encodes the one-loop potential for SU(3) and similarly for any SU(N) gauge group: 
the loop integral in \labelcref{eq:ModePotenial} is the same, and, hence, we simply use therein the mode potential from SU(2), namely $V_\textrm{mode} = V_\textrm{pert}/2$, as given by \labelcref{eq:VeffPert}.

If we simply add a mass term $m^2_A$ in the gluon propagator, the corresponding loop would feature a thermal suppression factor $\sim \exp{-m_A/T}$. 
Then, for $T\to 0$, the gluon loop would be exponentially suppressed, while the ghost loop would persist, thus reversing the overall sign of \labelcref{eq:VeffPert}. 
In this hypothetical case, the minimum would be $\varphi=1/2$ for $T\to 0$, and the Polyakov loop would vanish. 
Instead, without such a suppression, the gluon contribution would dominate for all temperatures and the confinement-deconfinement transition would not take place. 
In a nutshell, this is the confinement criterion put forth in \cite{Braun:2007bx}: \textit{Confinement in Landau-gauge QCD requires a gapped gluon propagator}. 
Evidently, one may mimic this confinement signature by simply adding a gluon mass, but this does not entail confinement: it lacks the dynamical mechanism underlying confinement.

%
\begin{figure*}[t]
\includegraphics[width=0.45\textwidth]{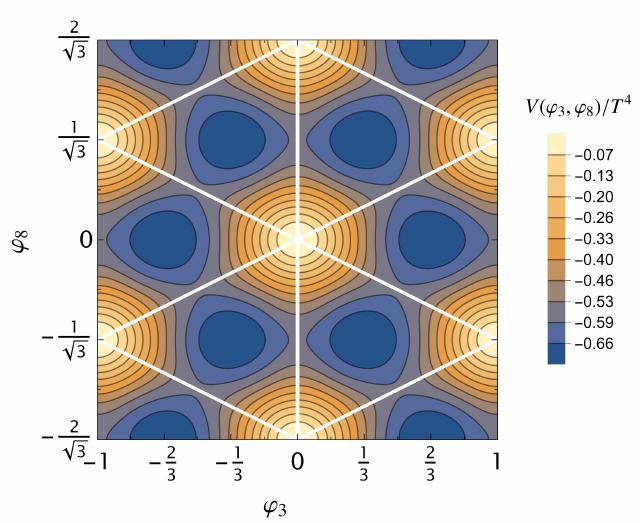}
\includegraphics[width=0.45\textwidth]{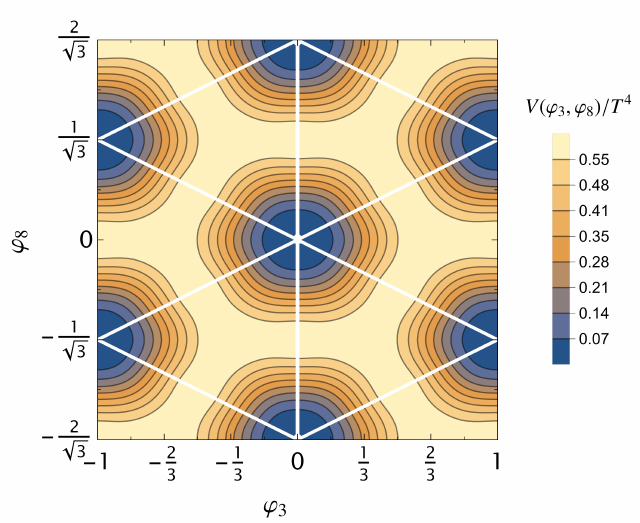}
\caption{Contour plot of the potential $V(\varphi_3,\varphi_8)$ in the confined (left, $T = 230$~MeV) and deconfined (right, \mbox{$T = 330$~MeV}) phases. The lines indicate the boundaries of the Weyl chambers (triangles). The centers 
of the Weyl chambers single out the center symmetric points of the Polyakov loop, and those at the vertices of the triangles have maximal breaking of center symmetry, $|L|=1$.  \hspace*{\fill}}
\label{fig:V_full}
\end{figure*}
%

The computation with the full propagators can be carried out analogously: 
we simply implement inside \labelcref{eq:ModePotenial} the substitution $G^{(0)}_i\to G_{i}$, with $G_{i}$ given by \labelcref{eq:GluonProp,eq:GhostProp}. 
Importantly, the location of the minima and maxima of the mode potentials $V_{A,c}$ does not change in this general case, as they are dictated by center symmetry. 
However, the shape of the potentials changes, making them analytically inaccessible.

\subsubsection{Critical temperature and the gluon mass gap}
\label{sec:Tconf}

It is left to determine the critical temperature. 
The qualitative analysis above illustrates that the confinement-deconfinement phase transition is triggered by the exponential suppression of the gluon contribution due to the presence of the gluon mass gap in the gluon propagator. 
We have performed the computation in SU(3), using the $T=0$ gluon and ghost propagators displayed in \Cref{fig:gluon_dressing_scales} and \Cref{fig:ghost_dressing}, respectively. 
The critical temperature $T_c\approx 275$\,MeV agrees quantitatively with that computed on the lattice.

The full effective potential $V_\textrm{eff}$, computed from the full Yang-Mills propagators, shows the pattern discussed in our SU(2) example with a simple massive gluon propagator: 
For small temperatures, the gluon part of the potential is exponentially suppressed and $V_\textrm{eff}$ approaches the ghost part of the potential with its global minus sign. 

In \Cref{fig:V_full} we show two contour plots of the potential $V_\textrm{eff}(\varphi_3,\varphi_8)$ for $T=230$\,MeV$< T_c$ (left plot) and \mbox{$T=330$\,MeV$> T_c$}. 
In both cases, one clearly sees the $Z_3$ center symmetry of the potential: \\[-2ex]

For $T>T_c$, the minima of the potential are the center broken points, leading to $|L|\to 1$. 
Due to center symmetry, the maxima are at the center-symmetric points with $L=0$. \\[-2ex]

For $T>T_c$, the minima become the maxima and vice versa, as already discussed in the simple example case. 
Hence, in this regime, the Polyakov loop vanishes. \\[-2ex]

\begin{figure}[t]
  \includegraphics[width=0.45\textwidth]{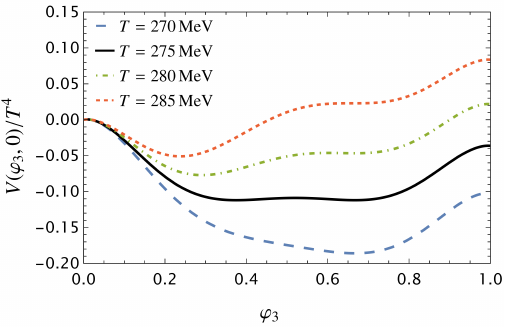}
  \caption{Polyakov potential $V(\varphi_3, \varphi_8)$ for $\varphi_3 \geq 0$, $\varphi_8 = 0$, and various temperatures close to the critical one, \mbox{$T_c = 275$}~MeV. \hspace*{\fill}}
  \label{fig:Vslices}
\end{figure}

If we bias the system towards real-valued Polakov loops, $L(\varphi)\in \mathbbm{R}$ by introducing an infinitesimal external current, we get $\langle \varphi_8\rangle =0$. 
The traced Polyakov loop of the expectation value is determined by the minimum of the potential $V_\textrm{eff}(\varphi_3)=V_\textrm{eff}(\varphi_3,\varphi_8=0)$, which is displayed in \Cref{fig:Vslices} for several temperatures around the phase transition. 
The first-order transition is clearly visible, as is the steepness of the order parameter $L(\langle \varphi\rangle)$ as a function of temperature.

\subsubsection{Infrared (in)sensitivity of the critical temperature}
\label{sec:IR-Tconf}

We finally turn to one of the main results of this Section, and discuss the sensitivity of the critical temperature $T_c$ on infrared deformations of the gluon propagator. 
The set of deformations used is shown in \Cref{fig:IR-DeformationsYM}, and their parametrisation is shown in \Cref{app:prop_vars}. 
This particular exercise consists of determining $T_c$ using all the gluon propagators shown in \Cref{fig:IR-DeformationsYM+QCD}.

The results of this procedure are shown in \Cref{fig:Tc-Deformations}: 
the critical temperature is independent of these deformations, as long as they only affect momenta approximately smaller than the infrared inflection point $p_{\textrm{in}}^- \approx 370$~MeV, see \Cref{tab:Inflection}. 
For deformations above this scale, $T_c$ starts to change. 
A basic understanding of this property is achieved by having a closer look at the loop expression \labelcref{eq:ModePotenial} for the mode potential. 
The spatial momentum integration can be rewritten in terms of an integral of the radial momentum $\sqrt{\vec p{}^2}$, leading to 
\begin{align} 
\int \frac{d^3 p}{(2 \pi)^3}\, G_i\bigl(p_0^2+\boldsymbol{p}^2\bigr) =\int \frac{d p}{2 \pi}\,\bigl[ p^2\, G_i\bigl(p_0^2+\boldsymbol{p}^2\bigr)\bigr]  \,, 
\label{eq:VeffDressing}
\end{align} 
with $p_0= 2 \pi T(n+\nu_j)$. 
At $p_0^2=0$, the expression in the square bracket is nothing but the propagator dressing $1/Z_i$ also shown in \Cref{app:prop_vars}: 
deformations below $p_{\textrm{in}}^-$ essentially leave no trace in the dressing. 
Moreover, the relevant $\varphi_3$-regime has $\varphi_3\gtrsim 0.2$, which translates into a shift of the Matsubara frequency by approximately 300\,MeV or more for temperatures around $T_c$. 
This reduces the relevance of lower momenta in the gluon (and ghost) propagators even further. 

In summary, the critical temperature $T_c$ is insensitive to the momentum dependence of the gluon propagator in the regime $p\lesssim p_{\textrm{in}}^-$. 
This reduced infrared sensitivity extends to other QCD observables that are obtained from gauge-fixed correlation functions by momentum integrals of their combinations. 
Indeed, the other observable considered in the present work, $f_\pi$, is also of this type.

%
\subsection{DCSB and the gluon mass gap}
\label{sec:ChiralGap} 

In this Section we explore the impact of the gluon mass gap on DCSB, by computing the pion decay constant, $f_\pi$, from the quark propagator in 2+1 flavour QCD, in the chiral limit; for state-of-the-art functional studies of the coupled system of the quark gap equation and the quark-gluon vertex, see \cite{Williams:2014iea, Mitter:2014wpa, Williams:2015cvx, Cyrol:2017ewj, Gao:2021wun,Aguilar:2024ciu, Fu:2025hcm}. 

The present analysis is based on \cite{Aguilar:2024ciu}. Specifically, we study the impact of the infrared deformations of \Cref{fig:IR-DeformationsYM+QCD} on $f_\pi$, see \Cref{fig:Observables-DeformationsYM+QCD}, as well as the constituent quark mass function $M_q(p)$, see \Cref{fig:M_vars}.

\subsubsection{Quark gap equation}
\label{sec:QuarkGap}

\begin{figure}[t]
	\centering
	\includegraphics[width=0.48\textwidth]{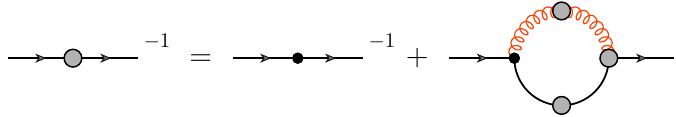} 
	\caption{ Quark propagator DSE. Full propagators are denoted by lines with grey circles, and the classical
vertices and propagators are indicated by small black circles. \hspace*{\fill} }
	\label{fig:gap}
\end{figure}

This analysis is based on the solution of the quark gap equation in Landau-gauge QCD. 
The quark propagator is parametrised as follows, 
\begin{subequations}
    \label{eq:QuarkProp}
\begin{align} 
G_{q\bar{q}}^{AB}(p) = \delta^{AB}G_{q\bar{q}}(p)\,,
\label{eq:Gqab}
\end{align} 
with $A,B=1,2,3$ are the color indices in the fundamental representation and 
\begin{align}
G_{q\bar{q}}(p) = \frac{1}{Z_q(p)[ i \slashed{p} + M_q(p) ] } \,.
\end{align}
\end{subequations}
The momentum evolution of $G_{q\bar{q}}(p)$ is determined by the quark gap equation, shown in \Cref{fig:gap}. In the Landau gauge and in the chiral limit, it reads 
\begin{subequations}
    \label{eq:DSE-Quark}
\begin{align}
G_{q\bar{q}}^{-1}(p) = i Z_2 \slashed{p} + \Sigma(p) \,, \label{eq:quark_gap}
\end{align}
where $Z_2$ is the quark wave-function renormalisation constant, and the renormalised quark self-energy, $\Sigma(p)$, is given by
\begin{align}\nonumber 
\Sigma(p) =&\, Z_1^f\, g_s\, C_f \! \int \!\! \frac{d^4 k}{(2 \pi)^4} G_A(k)\gamma_\mu \\[1ex] 
& \hspace{1.9cm}\times G_{q \bar{q}}(k+p) \,\Gamma^\bot_\mu (k+p,-p) \,, 
\label{eq:quark_self_energy}
\end{align}
\end{subequations}
where the contraction over color indices has been performed. 
In \labelcref{eq:quark_self_energy}, $g_s$ is the strong coupling and $C_f=4/3$ is the Casimir eigenvalue of the fundamental representation of SU(3). 
The vertex $\Gamma^\bot_\mu (k+p,-p)$ is the transversely projected quark-gluon vertex, and $Z_1^f$ its renormalisation constant. \Cref{eq:quark_self_energy} can then be projected on its Dirac structures to obtain a coupled system of DSEs for $Z_q(p)$ and $M_q(p)$. 
We fix $Z_2$ with a momentum-subtraction (MOM)-type renormalisation condition~\cite{Celmaster:1979km,Hasenfratz:1980kn,Braaten:1981dv,Athenodorou:2016oyh}, \ie $Z_q(p) = 1$ at $p^2 = \mu^2$.

The transversely projected quark-gluon vertex, $\Gamma^\bot_\mu$ contains eight independent tensor structures. 
In a self-consistent analysis, the corresponding form factors should be determined from their own dynamical equations~\cite{Mitter:2014wpa, Cyrol:2017ewj, Williams:2015cvx, Gao:2021wun, Aguilar:2024ciu}. 
As was shown in these works, apart from the classical tensor structure, proportional to $\gamma_\mu$, the inclusion of two further tensorial structures is required in order to achieve quantitatively accurate results for the quark propagator. 

The focus of the present analysis is the impact of infrared deformations of the gluon propagator on chiral observables. 
Therefore we consider a simplified setting, which captures the qualitative properties of DCSB: 
we drop the non-classical tensor, and include only the classical form factor $\lambda_1(k+p,-p)$, 
namely 
\begin{align}
\Gamma^\bot_\mu (k+p,-p) \to \Pi^\bot_{\mu\nu}(k)\, \gamma_\nu\, g_s\, \lambda_1(k+p,-p) \,. 
\label{eq:Gammamubot}
\end{align}
For the present analysis we use data from~\cite{Aguilar:2024ciu} 
for $\lambda_1(k+p,-p)$, together with the fit for the gluon propagator $G_A(k)$ shown as a black-dashed curve in \Cref{fig:IR-DeformationsQCD}. 
Moreover, we compensate for the lack of the additional tensor structures by tuning the coupling $\alpha_s$ such that the constituent quark mass at vanishing momentum takes the value \mbox{$M_q(0) = 350$~MeV}. 
This is achieved with the coupling 
\begin{align} 
\alpha_s = \frac{g_s^2}{4\pi} = 0.54\,,\qquad \mu =4.3\,\textrm{GeV}\,, 
\label{eq:RG-Gap}
\end{align}
at the renormalisation point with the RG-conditions \labelcref{eq:RG-Conditions} for the propagator dressings. \Cref{eq:RG-Gap} implies that $\alpha_s=\alpha_s(p^2=\mu^2)$, where $\alpha_s(p)$ is the running quark gluon coupling. 

The respective mass function $M_q(p)$ is shown in 
\Cref{fig:M_vars} as the continuous black line. 
Note that the constituent quark mass \mbox{$M_q(0) = 350$~MeV} is that of  2+1 flavour QCD, which is slightly higher than that in the chiral limit. 
However, for the present purpose the precise value is not of primary importance, as we are only interested in the relative impact of the infrared deformations of the gluon propagator. 
This concludes the discussion of the setup, including the computational benchmark result for the quark propagator.

\subsubsection{Infrared (in)sensitivity of the constituent quark mass}
\label{sec:IR-Mq}

In \Cref{sec:FullPolyakov} we have argued that the insensitivity of $T_c$ and further observables to infrared deformations of the gluon propagator can be explained by the appearance of the gluon dressing $1/Z_A$ in the loop integrals; see in particular \labelcref{eq:VeffDressing} and the related discussion. 
Now we apply this same argument to the gap equation \labelcref{eq:DSE-Quark}. 
In particular, we concentrate on the scalar part of the gap equation, $M_q(p)$, which reads
\begin{align}
M_q(p) = \frac{2\alpha_s}{\pi^2}Z_q^{-1}(p) \int\limits_0^\infty dk^2 \,\underbrace{k^2 G_A(k)}_{1/Z_A(k)} f(p,k) \,, \label{eq:quark_mass} 
\end{align}
where
\begin{align}
f(k,p) = \int\limits_0^\pi  d\theta \sin^2\theta\frac{M_q(u)\lambda_1(k+p,-p)}{Z_q(u)[ u^2 + M_q^2(u)]} \,,
\end{align}
with $u = k+p$, and the angle $\theta$ between $k$ and $p$. 

\Cref{eq:quark_mass} has the same structure as \labelcref{eq:VeffDressing}. 
The integrand is proportional to the dressing $1/Z_A(k)$ of the gluon propagator. 
The only difference is the factor $f(k,p)$, with a finite limit as $k\to 0$. 
In conclusion, the gap equation for $M_q(p)$ is effectively insensitive to variations of $G_A(p)$ that do not significantly change the gluon dressing function, that are deformations for momenta $p\lesssim p_{\textrm{in}}^-$, as discussed around \labelcref{eq:VeffDressing}. 
This argument is readily extended to the equation that controls the momentum evolution of the form factor $Z_q(p)$. 

\begin{figure}[t]
  \includegraphics[width=0.45\textwidth]{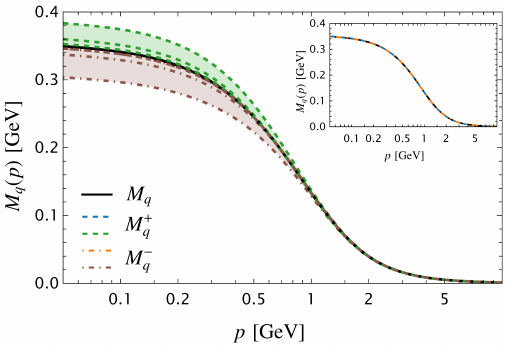}
  \caption{ Quark mass functions, $M_q(p)$ and $M_q^\pm(p)$, obtained with the gluon propagator, $G_A(p)$, and its variations, $G_A^{\pm}(p)$, shown in \Cref{fig:IR-DeformationsYM+QCD}. Curves and bands are colored according to the gluon propagators used as inputs. The inset shows only the $M_q^\pm(p)$ obtained when varying the gluon propagator below the infrared inflection point.\hspace*{\fill}}
  \label{fig:M_vars}
\end{figure}

Consequently, we expect that the quark propagator, and in particular the constituent quark mass function $M_q(p)$, is only sensitive to infrared deformations of momenta $p\gtrsim p_{\textrm{in}}^-$ in the 2+1 flavour gluon propagator in \Cref{fig:IR-DeformationsQCD}, depicted as green- and brown-shaded areas. 
The computational results for $M^\pm_q(p)$ are shown in \Cref{fig:M_vars}, and fully confirm our expectations. The colors correspond to those of the gluon propagators used as input, depicted in \Cref{fig:IR-DeformationsYM+QCD}. 
The only visible difference between $M_q(p)$ (black curve) and $M_q^\pm(p)$ originates from varying the gluon propagator beyond its infrared inflection point (green and brown bands and curves). 
If the results from these deformations are removed, the remaining curves (blue and orange) are indistinguishable from $M_q(p)$, as shown in the inlay of \Cref{fig:M_vars}. 

\subsubsection{Infrared (in)sensitivity of the pion decay constant}
\label{sec:IR-fpi}

The two dressing functions $Z_q$ and $M_q$ of the quark propagator are not observables themselves, as they are coefficients of a gauge-fixed correlation function. 

However, as discussed before, they enter 
into observables that are defined as momentum integrals of products or ratios of correlation functions. 
The most direct relation of $Z_q,M_q$ to a genuine observable is provided for the pion decay constant $f_\pi$. 
This relation is simplified further if using the Pagels-Stokar formula~\cite{Pagels:1979hd,Cornwall:1980zw},
\begin{align}
f_\pi^2 =&\, 4 N_c \!\! \int \!\! \frac{d^4 p}{(2 \pi)^4} \frac{\bar{Z}_2}{Z_q(p)}\frac{M_q(p)}{[p^2 + M^2_q(p)]^2}
 \nonumber\\[2ex]
&\, \hspace{2cm} \times \left[ M_q(p) - \frac{p^2}{2} M^\prime_q(p) \right] \,, \label{eq:fpi_PS}
\end{align}
where $M^\prime_q(p) = \partial_{p^2}M_q(p)$. 
The constant $\bar{Z}_2$ is identified with $Z_q(0)$ in the derivation of \labelcref{eq:fpi_PS} through the pion Bethe-Salpeter equation. 
Using the quark propagator from  \Cref{sec:QuarkGap} we obtain a pion decay constant of \mbox{$f_\pi = 93.2$~MeV}, which compares very well with $f_{\pi} \approx  93$~MeV for physical current masses. 
This proximity of values is expected, as we have tuned the constituent quark mass to the physical case and not to the chiral limit. 
In the latter case the constituent quark mass is slightly smaller, and so is the pion decay constant, \mbox{$f_{\pi,\chi} \approx 88$~MeV}, see e.g.,~\cite{FlavourLatticeAveragingGroup:2019iem} for chiral extrapolations of lattice results. 

Now we use \labelcref{eq:fpi_PS} to  study the impact of the infrared deformations of the gluon propagator on the pion decay constant: 
we compute the quark wave function $Z_q^\pm(p)$ and the constituent mass function $M_q^\pm(p)$ for these deformations, see \Cref{fig:M_vars}. 
Inserting these results in \labelcref{eq:fpi_PS}, we obtain the corresponding variations of the pion decay constant, denoted by $f^\pm_\pi$. 

From \Cref{fig:M_vars} we expect that only the deformations with $p\gtrsim p_{\textrm{in}}^-$ have a sizeable impact on the pion decay constant, since only the deformations in the green- and brown-shaded areas modify the quark propagator appreciably. 
This expectations is confirmed by the explicit computation: 
in \Cref{fig:fpi-Deformations}, we show the ratio $f_\pi^\pm/f_\pi$, between the pion decay constant obtained with the modified gluon propagator and that obtained with the lattice fit. 
Notably, $f_\pi^\pm/f_\pi$ is practically unity for variations of the gluon propagator below $390$~MeV, for which $f_\pi^\pm$ and $f_\pi$ deviate by less than $1\%$.

\section{The gluon mass gap}
\label{sec:GluonMassGap} 

\Cref{sec:Schwinger+massgap} and  \Cref{sec:GluonGapObservables} have shown that the gluon propagator incorporates a physical mass scale, the gluon mass gap $m_\textrm{gap}$, related to confinement. 
Moreover, important infrared observables, such as the confinement-deconfinement phase transition and $f_\pi$, directly depend on it. 
As discussed in the introduction of \Cref{sec:Schwinger+massgap}, similarly to the definition of the scale of chiral symmetry breaking, there is no unique definition of this mass gap. 
However, as we will explain in \Cref{sec:ScreeningMass}, a natural definition is that of the ``screening mass''. 
This identification also prompts us to put forth a simple, physically motivated fit, presented in \Cref{sec:ScreeningMassFits}.
This fit provides some insight into the relation between chiral symmetry breaking and confinement. 
First steps in this direction are taken in \Cref{sec:TfpimGap}, where we study the $m_\textrm{gap}$-dependence of \texorpdfstring{$T_c$}{Tconf} and \texorpdfstring{$f_\pi$}{fpi}.

\subsection{Screening mass}
\label{sec:ScreeningMass}

In the following we define $m_\textrm{gap}$ as a screening-type mass of the gluon propagator: \\[-2ex]

\textit{$m_\textrm{gap}$ is the real part of the location of the first (non-trivial) singularity in the complex plane, see \labelcref{eq:omegaGap}.}\\[-2ex] 

This choice is guided by two related aspects of infrared QCD:\\[-2ex]

\begin{figure}[t]
  \includegraphics[width=.49\textwidth]{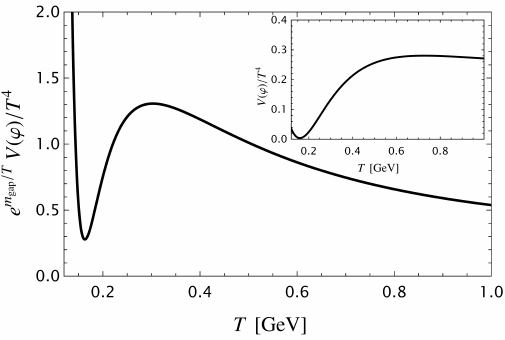}
  \caption{Gluon part of the Polyakov loop potential at the center-symmetry point $\varphi=1/2$, multiplied by $\exp\{m_\textrm{gap}/T\}$, with $m_\textrm{gap}$ from \Cref{tab:fit_params}.
  The asymptotic saturation of the result indicates that 
  the potential $V_A$ drops 
  as $\exp\{-m_\textrm{gap}/T\}$, see inlay.
\hspace*{\fill}}
  \label{fig:VA-Exp}
\end{figure}
%
$({\it 1})$ First, the computation of the transition temperature and similar finite temperature observables with the Matsubara formalism is tantamount to summing over the pole and cut contributions in the upper complex half plane. 
The Matsubara sums originate from the singularities induced by the thermal distribution. Using Cauchy's theorem, they can be 
obtained from a computation on the real frequency axis, encompassing physical singularities, such as pole and cut contributions. 
The latter include the contributions of the (upper half plane) singularities in the gluon propagator, as also discussed in ~\Cref{sec:FullPolyakov}: 
the details of the computation of $T_c$ unravel the fact that it is the first pole of the gluon propagator in the complex plane which provides the exponential suppression of the gluon loop for small temperatures. 
Moreover, for asymptotically small temperatures, the cut of the gluon propagator, induced by the ghost loop~\cite{Aguilar:2013vaa}, becomes visible. 
This is illustrated in \Cref{fig:VA-Exp}, where we show the gluon part of the Polyakov loop potential in $N_c=2$ Yang-Mills theory at the center-symmetric point $\varphi=1/2$, multiplied by $\exp\{ m_\textrm{gap}/T\}$. 
Here, $m_\textrm{gap}$ is provided by \Cref{tab:fit_params}: it is obtained by our determination of the screening mass from the physically motivated fit in \Cref{sec:ScreeningMassFits}.\\[2ex]

$({\it 2})$  Second, computations of the hadron resonance spectrum are performed with real-time bound state equations.
These equations require as inputs the QCD correlation functions, and, particularly, the gluon propagator, in the complex plane. 
The importance of the gluon propagator is highlighted in glueball computations, see e.g.,~the recent work \cite{Huber:2025kwy} and references therein, and these relations have already been exploited in \cite{Pawlowski:2022zhh}.  \\[-2ex]

The definition of $m_\textrm{gap}$ as the screening mass translates into a simple condition in terms of the gluon dressing $Z_A(p)$,    
\begin{subequations} 
\label{eq:omegaGap}
\begin{align} 
m_\textrm{gap} = \omega_\textrm{gap} \,,\qquad \omega_\textrm{gap} =  \left(1  + \imag \gamma_\textrm{gap}\right)\,m_\textrm{gap}\,,
\label{eq:eq:omegaGap1}
\end{align} 
with 
\begin{align} 
\omega_\textrm{gap} = \min_{\omega_s\neq 0}\{\textrm{Re}\,{\omega_s} \,|\, Z_A(-\omega_s^2) =0\} \,.
\label{eq:omegaGap2}
\end{align} 
\end{subequations}
For a physical resonance, $m_\textrm{gap}$ would 
correspond to the pole mass and $\gamma_\textrm{gap} m_\textrm{gap}$ would be the width. 
Moreover, if $\omega_\textrm{gap}$ is reconstructed from Euclidean data, the ambiguity of its real part, $m_\textrm{gap}$, is far smaller than that of $\gamma_\textrm{gap}$. 

The RGI property of $\omega_\textrm{gap}$, and hence that of $m_\textrm{gap}$, is readily deduced from the RG-equation of the gluon propagator, 
\begin{align} 
\left( \mu\frac{d}{d \mu }+2\,\gamma_A\right) G_A(p) =0\,, \qquad \mu\frac{d A_\mu} {d \mu } = \gamma_A \,A_\mu\,.
\label{eq:RGGA}
\end{align}
Here, $\mu$ is the RG-scale and $d/d\mu$ is the total derivative with respect to $\mu$. 
Evidently, \labelcref{eq:RGGA} implies  
\begin{align} 
\left( \mu\frac{d}{d \mu }-2\,\gamma_A\right) Z_A(p) =0\,, 
\label{eq:RGZA}
\end{align} 
and hence the zeros of $Z_A(p)$ are RGI, $\mu\, d Z_A(\omega_s)/ d\mu=0$. 
It is important to emphasise that \labelcref{eq:omegaGap} does not depend on the presence or absence of a spectral representation of the gluon, for respective discussions see e.g.,~\cite{Cyrol:2018xeq, Binosi:2019ecz, Fischer:2020xnb, Horak:2022myj, Loveridge:2022sih}. 

As mentioned before, with \labelcref{eq:omegaGap}, the gluon mass gap is defined similarly to a pole or screening mass of a physical particle. 
We hasten to emphasise, however, that the gluon does not define an asymptotic state that would be proportional to $A_\mu |\Omega\rangle$, where $|\Omega\rangle$ is the QCD vacuum.
Therefore, in order to avoid the misinterpretation of these similarities,  we have opted for the notion of a ``gluon mass gap'' rather than that of a ``gluon mass''.

\subsection{Fit of the gluon propagator using \texorpdfstring{$m_\textrm{gap}$}{mgap}}
\label{sec:ScreeningMassFits}

\begin{table*}[t]
\begingroup
\renewcommand{\arraystretch}{1.3}
\setlength{\tabcolsep}{6pt} 
\begin{tabular}{|c||c|c||c|c|c|c|c|c|c|}
	\hline
	$N_f$  & $m_\textrm{gap}[\textrm{MeV}]$& $\gamma_\textrm{gap}$ & $z_\textrm{sat}$ & $z_\textrm{peak}$& $z_\textrm{gh}$& $z_\textrm{uv}$ & $c_\textrm{uv} $ & 
     $c_+$  & 
     $c_-$ \\
	\hline
	$0$  & 
    686 & 
    0.585 & 
    0.245 & 
    0.670 & 
    0.036 & 
    0.853 & 
    0.326 &
    0.074  & 
     3.71 \\
    \hline 
    $2+1$  & 
    818 & 
    0.399 & 
    0.279 & 
    0.175 & 
    0.050 & 
    0.540 & 
    1.47  & 
    0.162 &
     3.70 \\
	  \hline
\end{tabular}
\caption{Parameters in  \labelcref{eq:GlobalFitGluon} for the gluon propagator in Yang-Mills theory ($N_f=0$), and the $N_f=2+1$ flavour gluon propagator. The fits in comparison to the functional and lattice data are shown in  \Cref{fig:gluon_dressing_scales}. \hspace*{\fill}}
\label{tab:fit_params}
\endgroup
\end{table*}

In this Section,  we propose a simple global fit for the gluon propagator, where the scale setting is implemented through 
$m_\textrm{gap}$ rather than $\Lambda_\textrm{QCD}$~\cite{Zafeiropoulos:2019flq}: 
this identification uses the fact that they are directly related and we cannot change them independently. In \Cref{app:ChangeMassgap} we discuss this in more detail. 
Moreover, all further fitting parameters admit a direct ``physical'' interpretation within the three regimes that were identified in \Cref{sec:Schwinger+massgap}:\\[-2ex] 

$(uv)$ the perturbative ultraviolet regime, \\[-2ex] 

$({\it sc})$ the strongly correlated Schwinger regime, \\[-2ex] 

$({\it ir})$ the deep infrared regime.  \\[-2ex] 

\noindent Two further parameters cover the interfaces between the three regimes, \\[-2ex] 

$({\it int})$ the two interfaces between (\textit{uv,sc,ir}). \\[-2ex] 

\noindent In fact, the corresponding fitting parameters are fixed {\it locally} within the respective regimes.

\subsubsection{General structure of the fit}
\label{sec:thefit}

For decoupling-type of gluon propagators, the proposed fit takes the form 
\begin{subequations} 
\label{eq:GlobalFitGluon}
\begin{align} 
Z_{A,\textrm{fit}}(p) =Z_{A,\textrm{ir}}(x) +Z_{A,\textrm{uv}}(x) \,,
\label{eq:GlobalFitGluonGA}
\end{align}
with
\begin{align}
Z_{A,\textrm{ir}}(x) =\frac{z_\textrm{sat}/x-z_\textrm{peak}- z_\textrm{gh} \log \left( 1+\frac{1}{c_- x}\right)} {(1+c_+ x)^2} \,,
\label{eq:ZAir}
\end{align}
and
\begin{align}
Z_{A,\textrm{uv}}(x) = z_\textrm{uv}\bigl[1 + c_\textrm{uv}\log \left(1+ c_+\,x\right)\bigr]^\gamma\,. 
\label{eq:ZAuv}
\end{align}
\end{subequations}
The dimensionless parameter $x$ denotes the only available kinematic variable, corresponding to the physical momentum squared measured in units of $m_\textrm{gap}$, namely  
\begin{align} 
x(p)=\frac{p^2}{m_\textrm{gap}^2}\,. 
\label{eq:DimlessMomenta}
\end{align}
Finally, the parameter $\gamma$ captures the one-loop anomalous dimension, 
\begin{align}
\gamma=\frac{13- \frac43\, N_f}{22- \frac43  \,N_f}\,,
\label{eq:gamma}
\end{align} 
while the exponentiated form of the square bracket comes from a one-loop resummation. 

We close this part with a brief discussion of the scales in \labelcref{eq:GlobalFitGluon}, as the gluon dressing has a dependence on the mass gap $m_\textrm{gap}$, on $\Lambda_\textrm{QCD}$, and on the RG-scale $\mu$. 
As mentioned before \labelcref{eq:GlobalFitGluon}, we have used that $m_\textrm{gap}, \Lambda_\textrm{QCD}$ are both avatars of the dynamically generated mass scale of QCD, namely its physical mass gap, to wit, 
\begin{align} 
\Lambda_\textrm{QCD} \propto m_\textrm{gap}\,. 
\label{eq:mGapLambdaQCD}
\end{align}
In \labelcref{eq:GlobalFitGluon} this relation is used and only $m_\textrm{gap}$ is present. 
This entails specifically, that in Yang-Mills theory a change of $m_\textrm{gap}$ amounts to no change of the physics. 
In turn, changing one of the two scales while keeping the other fixed, is a physics change and hence we do not describe Yang-Mill theory any more. On the other hand, in QCD we can use a change of $m_\textrm{gap}$ with \labelcref{eq:mGapLambdaQCD} for changing the scales of the confinement dynamics relative to that of the chiral dynamics: 
qualitatively, this emulates changing the color and flavour number which has been shown to change the ratio $m_\textrm{gap}/m_\chi$, see \cite{Goertz:2024dnz}. 
This will be discussed further in \Cref{sec:TfpimGap}. 

The fit $Z_{A,\textrm{fit}}(p)$ in \labelcref{eq:GlobalFitGluon} carries no explicit dependence on the RG-scale $\mu$ and uses the RG-condition \labelcref{eq:RG-Conditions} at $\mu=4.3$\,GeV with 
\begin{align}
Z_{A,\textrm{fit}}\bigl(x(\mu)\bigr)=1, \qquad x(\mu)= \frac{\mu^2}{m_\textrm{gap}^2}\,.
\label{eq:RGcondition-x}
\end{align} 
Accordingly, a change of $m_\textrm{gap}\to \lambda\, m_\textrm{gap}$ at fixed fitting parameters amounts to a change of the RG-scale with $\mu\to \lambda\mu$. 
Moreover, the RG-equation \labelcref{eq:RGZA} for $Z_A$ entails that the fitting parameters are RG-scale dependent. 
This is discussed at the example of the UV-part $Z_{A,\textrm{uv}}$ in \Cref{app:ChangeMassgap} where we also discuss the relation of \labelcref{eq:GlobalFitGluon} to common fits in the literature.

\subsubsection{Physics \& determination of the fitting parameters}
\label{sec:FitPhysics-Extraction}

The gluon mass gap $m_\textrm{gap}$ is the only mass scale in \labelcref{eq:GlobalFitGluon}, taking over the rôle of $\Lambda_\textrm{QCD}$ in standard fits used in the literature~\cite{Cornwall:1981zr,Aguilar:2006gr,Aguilar:2010gm,Aguilar:2010cn,Gao:2021wun}. 
It only enters via $x$ defined in \labelcref{eq:DimlessMomenta}. 
Note also that $m_\textrm{sat}^2=G^{-1}_{A,\textrm{fit}}(0)= z_\textrm{sat} p^2/x$ is the saturation mass of the gluon propagator and hence $z_\textrm{sat}$ is the ratio of saturation mass and gluon mass gap squared. 

The other parameters in \labelcref{eq:GlobalFitGluon} can be `collected' into three sets belonging to the regimes ${(\it uv)}$ ${(\it sc)}$, and ${(\it ir)}$, 
\begin{align} 
{(\it uv)}:\ z_\textrm{uv}\,,\ c_\textrm{uv} \,,\quad  {(\it sc)}:\ z_\textrm{peak}\,, \quad {(\it ir)}:\ z_\textrm{sat}\,, z_\textrm{gh}   \,. 
\label{eq:3RegimesParameters}
\end{align} 
The last two parameters adjust the two interfaces between \textit{(uv,sc,ir)}. 
In particular, they accommodate the ``decoupling'' of the asymptotic scalings. 
\begin{align} 
(\textit{int}):\  c_+\,,\,c_-\,.
\end{align} 
For $c_+ x\gg 1$ the infrared dressing $Z_{A,\textrm{ir}}$ decouples, while for $c_- x\gg 1$ the infrared logarithm from the ghost loop decouples. 

The respective results for all parameters are collected in \Cref{tab:fit_params}, as well as the ``width'' parameter $\gamma_\textrm{gap}$ in \labelcref{eq:omegaGap}. 

We proceed with the determination of the fitting parameters. At first, one might be tempted to use \labelcref{eq:GlobalFitGluon} as a global fit, and fix the parameters through standard $\chi^2$-minimisation. 
However, such a procedure ignores the fact that all parameters describe specific dynamical aspects in one of the three regimes, ${(\it uv)}$, ${(\it sc)}$, and ${(\it ir)}$. 
Therefore, they should be fitted within these regimes rather than globally, in order to best accommodate the corresponding physics.

Note that the description of the interfaces between regimes is assigned to only two of the parameters, namely 
$c_\pm$. 
This minimalistic choice is prompted by the fact that gluon and matter correlation functions show rather rapid transitions between regimes. 
Hence, the corresponding interfaces are rather limited, suggesting that the parameters $c_\pm$ should suffice for a reasonable description of the transitions. 
Our results confirm this expectation: 
the fitting quality around the interfaces is only slightly inferior to the rest of the regimes. 

For the determination of the fitting  parameters from the numerical data, we rewrite the fit \labelcref{eq:GlobalFitGluon} in physical units:  
\begin{align}
x = \eta\, \bar x\,, \qquad \eta = \left( \frac{m_\textrm{gap}^2}{1 \textrm{GeV}^2}\right)^{-1} \to \bar x =p^2[\textrm{GeV}^2]\,. 
\end{align} 
Then, after rescaling all parameters in $Z_{A,\textrm{fit}}$ by $\eta$, we arrive at 
\begin{align} 
Z_{A,\textrm{fit}}(x) = \bar Z_{A,\textrm{fit}}(\bar x)\,.
\label{eq:barZA}
\end{align} 
$\bar Z_{A,\textrm{fit}}(\bar x)$ has the form
\labelcref{eq:GlobalFitGluon} with the parameters $\bar c,\bar z$, 
\begin{align}
 \bar c_\pm = \eta\, c_\pm\,,\qquad  \bar z_\textrm{sat} = \frac{z_\textrm{sat}}{\eta}\,, 
    \label{eq:barParameters1}
\end{align}
together with the invariant parameters 
\begin{align}
\bar c_\textrm{uv} = c_\textrm{uv}\,,\qquad \bar z_i =z_i\,,\quad i=\textrm{peak}, \textrm{gh} , \textrm{uv}\,.
    \label{eq:barParameters}
\end{align}
The parameters $\bar c\,,\,\bar z$ can be determined directly from gluon propagator data in their respective regimes ${(\it uv)}$ ${(\it sc)}$, and ${(\it ir)}$, introduced in \labelcref{eq:3RegimesParameters}: \\[-2ex]

\textit{UV regime} ${(\it uv)}$ : The two UV parameters capture the perturbative regime and are adjusted there, according to  
\begin{align}
\bar z_\textrm{uv}\,,\ \bar c_\textrm{uv}: \ \lim_{\bar x\to \infty}  Z_{A,\textrm{fit}}(p) = \lim_{x\to \infty} Z_\textrm{data}(p)\,.
\label{eq:Fix-UV}
\end{align}
In this limit, only the part $\bar Z_{A,\textrm{uv}}(\bar x)$ survives, with \mbox{$\bar Z_{A,\textrm{uv}}(\bar x) \to \bar z_\textrm{uv}\bigl[1 + \bar c_\textrm{uv}\log \left(x\right)\bigr]^\gamma$}. 

The validity regime of fits akin to that of \labelcref{eq:Fix-UV} has been studied in \cite{Gao:2021wun}, for 2+1 flavour QCD. 
As was shown there at the level of the quark-gluon running coupling, the momentum range compatible with two-loop perturbation theory extends to $p\approx 3$\,GeV. 
This is compatible with the location of the ultraviolet inflection point $p_{\textrm{in}}^+ \approx 1.6$~GeV, defined in \labelcref{eq:InflectionUV} as the approximate ultraviolet onset scale of the dynamics regime of the Schwinger mechanism. 
Similar considerations apply to Yang-Mills theory. 

\textit{Schwinger regime} ${(\it sc)}$: 
In this regime we have a single fitting parameter, $\bar z_\textrm{peak}$, whose value controls the peak height and position of the dressing function (red point in \Cref{fig:gluon_dressing_scales}).
Specifically, 
\begin{align} 
\bar z_\textrm{peak}:\ Z_{A,\textrm{fit}}(p_\textrm{peak})=Z_\textrm{data}(p_\textrm{peak})\,,
\label{eq:Fix-Schwinger}
\end{align}
In our procedure, it is determined through a global $\chi^2$-fit in the momentum interval between the two inflection points, $p_\textrm{in}^\pm$, as the regime is dominated by the emergence of the mass gap, and the subsequent decoupling of the dynamics. \\[-2ex]

\textit{Deep IR regime} ${(\it sc)}$: The infrared parameters are fixed through
\begin{align}\nonumber 
\bar z_\textrm{sat}:& \ G_{A,\textrm{fit}}(0) = G_\textrm{data}(0)\,, \\[2ex] 
\ \bar z_\textrm{gh}:&  \lim_{\bar x\to 0}\frac{\partial_{\bar x} G^{-1}_{A,\textrm{fit}}(p)}{\log \bar x} = \lim_{\bar x\to 0}\frac{\partial_{\bar x} G^{-1}_\textrm{data}(p)}{\log \bar x}\,.
\label{eq:Fix-IR}
\end{align}
The first condition in \labelcref{eq:Fix-IR} fixes the saturation mass squared, $m^2_\textrm{sat}$, with that from the gluon propagator data, namely $G_{A,\textrm{data}}(0)=m_\textrm{sat}^{-2}$. 
The second condition in \labelcref{eq:Fix-IR} adjusts the subleading logarithmic running of the gluon propagator, which originates from the massless ghost loop \cite{Aguilar:2013vaa}. 
To date, this deep infrared behaviour is only accessible within functional approaches; it cannot be extracted from the lattice data, as the respective lattice volumes are still too small. 
We have determined it from the ghost loop of the gluon DSE. 
Since the ghost propagator and ghost-gluon vertex are quite insensitive to the matter sector~\cite{Ayala:2012pb}, we use Yang-Mills results, namely the fit in \Cref{app:FitGhost} for the ghost dressing, and the ghost-gluon vertex from \cite{Ferreira:2023fva}, for the determination in both the $N_f = 0$ and $N_f = 2 + 1$ cases; the numerical difference in the values of $z_\textrm{gh}$ in \Cref{tab:fit_params} results from the different values of the strong coupling, $g_s$, in each case.  \\[-2ex]

\textit{Interfaces \textit{(int)}:} The final adjustment of the global fit is done with the parameters $\bar c_\pm$ that control the interfaces between the different regimes. They are determined together with $z_\textrm{peak}$  through a $\chi^2$-fit for momenta between the two inflection points $p_\textrm{in}^\pm$.

\subsubsection{Gluon mass gap and scale invariance}
\label{sec:FixGluonMassGap}

The above procedure fixes all parameters in $\bar Z_{A,\textrm{fit}}$. 
The mass gap, $m_\textrm{gap}$, and the  dimensionless ``width'' $,\gamma_\textrm{gap}$, are obtained by solving the (complex) pole condition \labelcref{eq:omegaGap},  
\begin{align} 
\bar Z_{A,\textrm{fit}}(-\bar\omega_s^2)=Z_{A,\textrm{ir}}(-\bar\omega_s^2)+Z_{A,\textrm{uv}}(-\bar\omega_s^2)=0\,, 
\label{eq:barZA=0} 
\end{align}
with $\bar \omega^2_s = (1 + \imag \gamma_\textrm{gap})^2/\eta$. \Cref{eq:barZA=0} fixes $\eta$ and $\gamma_\textrm{gap}$. 

The parameters $\bar c,\bar z$ and rescaling parameter $\eta$ allow us to determine $c,z$ by using \labelcref{eq:barParameters1,eq:barParameters,eq:mgapPhys}. 
The mass gap is simply given by  
\begin{align} 
m_\textrm{gap}[\textrm{GeV}]=\frac{1}{\sqrt{\eta}}\,. 
\label{eq:mgapPhys}
\end{align} 
The results for $m_\textrm{gap}, \gamma_\textrm{gap}$ for Yang-Mills theory and 2+1 flavour QCD are summarised in \Cref{tab:fit_params}, together with the other fitting parameters $c,z$. 
As discussed around \labelcref{eq:RGcondition-x}, the fit defines the RG-scale relative to the gluon mass gap, and we infer from \Cref{tab:fit_params} that 
\begin{align}
\mu_\textrm{\tiny{YM}}^2 \approx \, 39.29 \left(m^\textrm{(\tiny{YM})}_\textrm{gap}\right)^2\!\!,\quad  \mu_\textrm{\tiny{2+1}}^2 \approx \, 27.63\left(m^\textrm{(2+1)}_\textrm{gap}\right)^2\!\!,
\label{eq:DimlessRGscales}
\end{align}
for Yang-Mills theory ($\mu_\textrm{\tiny{YM}}$) and 2+1 flavour QCD ($\mu_\textrm{\tiny{2+1}}$) respectively. This concludes the discussion of the fit \labelcref{eq:GlobalFitGluon} for the gluon dressing $Z_A$. 

Note that, in Yang-Mills theory, this fit takes into account the (classical) conformal invariance of the theory: the quantum theory only depends on a single mass scale. 
Commonly, this scale is saturated by $\Lambda_\textrm{QCD}$, which may be extracted from the perturbative $\beta$-function. Nonetheless, we consider that the unique scale of the theory is best represented by the gluon mass gap,  generated by the Schwinger mechanism, and our fit makes this physical choice manifest. 
A fully self-consistent description may be given by using this scale-setting in the fits of all correlation functions. 
Then, a change of the gluon mass gap in the Yang-Mills gluon propagator does not change the theory at all, as dimensionless functions only depend on the $x$ defined in \labelcref{eq:DimlessMomenta}.  

In QCD the situation is different, as it features two dynamical infrared scales: the confinement scale, $m_\textrm{gap}$, 
and the DCSB scale, $m_\chi$; 
related observables/order parameters are $T_c$ and $f_\pi$. 
Therefore, in contradistinction to pure Yang-Mills, the variation of $m_\textrm{gap}$, while keeping the chiral scales fixed, changes the dynamics of the theory. 
Such a variation mimics changing colors and flavours (or current quark masses), and can be used for investigating the interrelation of confinement and chiral dynamics. 
This interrelation is relevant for the exploration of the phase structure of many flavour and color QCD, for a recent functional study, see \cite{Goertz:2024dnz}. 

We expect that variations of the fitting parameters in  \labelcref{eq:GlobalFitGluon}, and in particular of $m_\textrm{gap}$ relative to other 
intrinsic scales, should provide valuable insights on the intertwined infrared dynamics of QCD. 
A first attempt in this direction will be presented in the next subsection.

\subsection{Dependence of \texorpdfstring{$T_c$}{Tconf} and \texorpdfstring{$f_\pi$}{fpi} on the gluon mass gap}
\label{sec:TfpimGap}

In this final subsection we comment on the dependence of the two observables under investigation, $T_c$ and $f_\pi$, on the gluon mass gap. 
For this analysis we use the fits \labelcref{eq:GlobalFitGluon} with \Cref{tab:fit_params} for the gluon propagator in Yang-Mills theory and 2+1 flavour QCD. 

We first consider the confinement-deconfinement temperature. 
We have shown in \Cref{sec:ScreeningMassFits} that the gluon propagator can be written in terms of the dimensionless momenta $p^2/m_\textrm{gap}^2$, see 
\labelcref{eq:DimlessMomenta}, together with fitting parameters that do not depend on $m^2_\textrm{gap}$. 
Consequently, the critical temperature is simply given by 
\begin{align} 
T_c \approx c_\textrm{conf} \, m_\textrm{gap}\,, 
\label{eq:LinTconfMgap}
\end{align} 
where the value of $c_\textrm{conf}$ depends on the values of the other fitting parameters in \Cref{tab:fit_params}. 
We expect a subleading $m_\textrm{gap}$-dependence of $c_\textrm{conf}$, if the full dynamics is considered. 
In addition, corrections due to the implicit $m_\textrm{gap}$-dependence of the ghost must also be considered. 

Moving to 2+1 flavour QCD, and in particular to chiral observables such as $f_\pi$, the situation is considerably more complicated, due to the presence of two fundamental mass scales, that of DSCB, $m_\chi$, and that of confinement, related to $m_\textrm{gap}$. 
There are, however, two limiting cases that may be analysed qualitatively. 

For this analysis we use the fact that the fit \labelcref{eq:GlobalFitGluon} with its sole dependence on $x$ leads to \labelcref{eq:mGapLambdaQCD,eq:DimlessRGscales}. 
Hence, changing $m_\textrm{gap}$ with 
\begin{align} 
m_\textrm{gap}(\lambda) = \lambda\, m^\textrm{(2+1)}_\textrm{gap}\,,
\label{eq:lambda-mGap}
\end{align}
is tantamount to changing $ \Lambda_\textrm{QCD}$ and the RG-scale relative to the chiral symmetry breaking scale $m_\chi$;  
here, $m^\textrm{(2+1)}_\textrm{gap}=818$\,MeV from \Cref{tab:fit_params}. 
This implies an RG-scale 
\begin{align} 
\mu_{2+1}(\lambda) = \lambda\, \mu_{2+1}\,,\qquad \mu_{2+1}=4.3\,\textrm{GeV}\,,
\label{eq:lambda-mu}
\end{align} 
which follows from \labelcref{eq:DimlessRGscales}. 

Let us first consider the limit $m_\textrm{gap}/m_\chi\to \infty$. 
Heuristically, an increasing value of $m_\textrm{gap}$ 
leads to the gradual suppression of the gluon propagator, eventually reducing the overall interaction strength below the required critical size. 
Computationally, for $\lambda >1$, the gap equation in \labelcref{eq:quark_mass} is solved with
\begin{align} 
\alpha_s \to \alpha_s(\lambda \,\mu_{2+1})\,,
\label{eq:alphas-lambda}
\end{align}
with the running quark-gluon coupling $\alpha_s(p)$ in QCD, i.e.,~that used in \Cref{sec:ChiralGap}. 
With increasing value of 
$m_\textrm{gap}(\lambda)$, the input coupling decreases logarithmically. 
As the gap equation is formulated in $x$, the decrease of $\alpha_s$ is the only change. 
Accordingly, the constituent quark mass drops, and so does $f_\pi$. 
There is a critical value, $m_\textrm{gap}(\lambda^*)=m_\textrm{gap}^*$, beyond which DCSB is absent, with $M_q(p) \equiv 0$ in the chiral limit considered here. We are led to 
\begin{align} 
f_\pi(m_\textrm{gap}>m_\textrm{gap}^*)=0\,.
\label{eq:NoPion}
\end{align}
Now we consider the opposite limit with $m_\textrm{gap}/m_\chi\to 0$. Note that we do not encounter a Landau pole, as the present procedure keeps the ratio $\Lambda_\textrm{QCD}/m_\textrm{gap}$ fixed. Instead, if we take $m_\textrm{gap}\to 0$  at fixed $\Lambda_\textrm{QCD}$, we encounter a Landau pole, see \Cref{app:ChangeMassgap}.  

For fixed ratio $\Lambda_\textrm{QCD}/m_\textrm{gap}$, the Schwinger regime and the deep infrared regime are at far lower scales than that of DCSB for $m_\textrm{gap}/m_\chi\to 0$. 
At momenta $p\gtrsim m_\chi$, the gluon propagator is perturbative, resembling a photon-type propagator $\sim 1/p^2$, accompanied by the gluonic tail associated with asymptotic freedom. 

In this case, $f_\pi$ saturates at a value that we denote by $f_\pi^0$: the saturation is caused because the generation of a constituent quark mass switches off the infrared dynamics in the quark gap equation below the respective scale. 
Thus, we are led to
\begin{align} 
f_\pi\bigl(m_\textrm{gap}/m_\chi\to 0\bigr) = f_\pi^0\,.
\label{eq:fpiPhoton}
\end{align} 
Computationally, this limit comes with a subtlety: while we would like to use \labelcref{eq:alphas-lambda} with the physical quark-gluon coupling used in \Cref{sec:ChiralGap},  
this is a good approximation only in the regime where $\alpha_s$ runs perturbatively. 
Taking the inflection point $p_\textrm{in}^+$ as an approximate boundary of this regime, we are bound by $\lambda\approx 0.3$. 
For smaller $\lambda$ one may consider RG-scales that do not satisfy \labelcref{eq:lambda-mu}. 
This requires an explicit $\mu$-dependence of the fit, which is discussed in \Cref{app:ChangeMassgap}. 
In any case, the input coupling $\alpha_s(\lambda \mu_{2+1})$ is rising with decreasing $\lambda$. Accordingly, $f_\pi$ is a non-trivial and monotonically decreasing function of $m_\textrm{2+1}(\lambda)$. 

A reliable study of the related dynamics requires detailed numerical computations, along the lines of \cite{Goertz:2024dnz}. 
Such an analysis is important for further investigations of the phase structure of many flavour QCD, and especially a possible new phase characterised by the locking of confinement and DCSB, uncovered in \cite{Goertz:2024dnz}. 
However, a detailed computation lies beyond the scope of the present work, and shall be considered elsewhere.

\section{Conclusion}
\label{sec:Conclusion}

In this work we have discussed the physics of the gluon mass gap, $m_\textrm{gap}$, 
both in pure Yang-Mills theories and 
2+1 flavour QCD. 
We have argued that it is well-defined as a screening mass of the gluon propagator, even though its value is difficult to determine precisely. 

In addition, 
we have shown that important observables in infrared QCD are directly sensitive to the size of $m_\textrm{gap}$, but 
are insensitive to the momentum dependence of the gluon propagator below the decoupling scale induced by $m_\textrm{gap}$. 
Specifically, we have studied the $m_\textrm{gap}$-dependence of the critical temperature $T_c$ of the confinement-deconfinement phase transition, and the pion decay constant $f_\pi$. Importantly, these relations elevate $m_\textrm{gap}$ to an observable itself, even though it is only defined as an RGI property of a gauge-fixed correlation function. 
This property is similar to that of physical pole and screening masses that can be extracted from gauge-fixed correlation functions, if the latter have an overlap with the respective physical states. This motivated us to define the gluon mass gap as the screening mass of the gluon propagator. 

Our analysis led us also to a physically motivated fit for the gluon propagator, whose parameters are related to both ultraviolet and infrared properties of QCD, and include the mass gap. 
We have used this fit for the study of the $m_\textrm{gap}$-dependence of $T_c,f_\pi$ mentioned above, as well as in a preliminary analysis of the interplay between confinement and chiral symmetry breaking. 
We envisage the extension of this study with further observables, in order to achieve a comprehensive understanding of the intertwined dynamics. We hope to report on the respective results in the near future. 

Finally, an alternative approach to   
the dynamical generation of the gluon mass gap is the quartet mechanism \cite{Kugo:1979gm}. Similarly to the Schwinger mechanism, it proceeds through the formation of massless bound state poles in the ghost-gluon vertex \cite{Alkofer:2011pe}. 
It would be interesting to explore possible relations between the two mechanisms, in an attempt to unify both approaches. 

\section*{Acknowledgments}
\label{sec:Acknow}

We thank Reinhard Alkofer and Jan Horak for discussions and work on related subjects. M.N.F.~is supported by the National Natural Science Foundation of China (grants 12135007 and W2433021). 
The work of J.P.~is funded by the Spanish MICINN grants PID2020-113334GB-I00 and PID2023-151418NB-I00, the Generalitat Valenciana grant CIPROM/2022/66, and CEX2023-001292-S by MCIU/AEI. 
J.P.~is also supported in part by the EMMI visiting grant of the ExtreMe Matter Institute EMMI at the GSI, Darmstadt, Germany. 
J.M.P.~is funded by the Deutsche Forschungsgemeinschaft (DFG, German Research Foundation) under Germany’s Excellence Strategy EXC 2181/1 - 390900948 (the Heidelberg STRUCTURES Excellence Cluster) and the Collaborative Research Centre SFB 1225 - 273811115 (ISOQUANT).

\appendix

%
\section{Varying the gluon propagator}
\label{app:prop_vars}

In order to assess the (in)sensitivity of $T_c$, $M_q(p)$, and $f_\pi$, to the infrared behaviour of the gluon propagator, we deform $G_A$ in the deep infrared, either enhancing or suppressing it, without affecting its form for momenta $p$ larger than a chosen point, $\nu$.

Specifically, taking $G_A(p)$ to be the fit given by \labelcref{eq:GlobalFitGluon}, with the parameters in \Cref{tab:fit_params} for either $N_f = 0$ or $N_f = 2+1$, we construct variations given by
\begin{align} 
G_A(p) \to G_A^{\pm}(p) =&\, G_{A,\textrm{fit}}(p) \pm \delta(p) \,, \nonumber\\
\delta(p) =&\, \delta_0\exp[- (p/\nu)^\kappa ] \,. 
\label{eq:GA_vars}
\end{align}
For Euclidean $p > 0$, the term $\delta(p)$ can be seen as a smooth step function, affecting the propagator for momenta $p < \nu$, and vanishing quickly for $p > \nu$, where $G_A^{\pm}(p)$ reduces to $G_A(p)$. 
The exponent $\kappa$ controls how quickly $\delta(p)\to 0$ for $p >\nu$, while $\delta_0$ determines the size of the variation at the origin. 

In \Cref{fig:IR-DeformationsYM+QCD}, we show the gluon dressing function, $Z_A(p)$, resulting from the above variations for both the Yang-Mills theory (\Cref{fig:IR-DeformationsYM}) and 2+1 flavour QCD (\Cref{fig:IR-DeformationsQCD}), with the corresponding propagators displayed in the inlays. 
In both cases, we employ \mbox{$\delta_0 = 3.3$~GeV$^{-2}$}, $\kappa = 3$, and $\nu\in[0.1,0.5]$~GeV. In addition, 
in the same figure, we show as black continuous lines the original fits for $G_A(p)$, given by \Cref{eq:GlobalFitGluon}, the positions of the inflection points $p_{\textrm{in}}^-$ and $p_{\textrm{in}}^+$ (blue points), and of the maximum of $1/Z_A(p)$ (red point). 
The variations $G_A^+$ and $G_A^-$ are denoted by dashed and dot-dashed lines, respectively, and are colored differently for the cases where $\nu$ is larger or smaller than the infrared inflection point, $p_{\textrm{in}}^-$. 
Note that, at the level of $1/Z_A(p)$, the variations with $\nu < p_{\textrm{in}}^-$ are practically invisible.

%
\section{Ghost propagator}
\label{app:FitGhost}

\begin{figure}[t]
  \includegraphics[width=\linewidth]{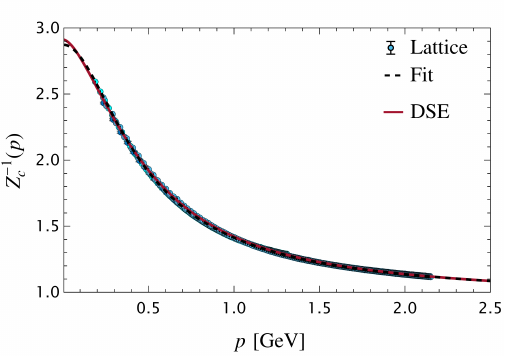}
  \vspace{-0.5cm}
  \caption{Dressing $1/Z_c(p)$ of the ghost propagator. 
  We show lattice data~\cite{Boucaud:2018xup} (data points), DSE data~\cite{Aguilar:2021okw} (red curve), and a fit (dashed black line) obtained with the lattice data and the perturbative UV behaviour.\hspace*{\fill}}
  \label{fig:ghost_dressing}
\end{figure}

In \Cref{fig:ghost_dressing} we show lattice data~\cite{Boucaud:2018xup} (data points) and DSE data~\cite{Aguilar:2024ciu} (red continuous line) for the inverse ghost dressing function, $Z_c(p)$. 
We also included a combined fit to the lattice data and the perturbative ultraviolet behaviour (dashed black line), 
\begin{align}
Z_{c,\textrm{fit}}(p)= Z_{c,\textrm{ir}}(p) + Z_{c,\textrm{uv}}(p) \,, \label{eq:GhostFit}
\end{align}
where
\begin{align}
Z_{c,\textrm{ir}}(p) = - \frac{z_\textrm{c,ir}}{1 + d_\textrm{ir}p^2} \,,
\end{align}
and
\begin{align}
Z_{c,\textrm{uv}}(p) = z_\textrm{c,uv}\left[ 1 + b_\textrm{uv}\log( 1 + d_\textrm{uv}p^2 ) \right]^{9/44}  \,,
\end{align} 
with the fitting parameters in \Cref{tab_GhostFitParameters}. As in the case of the gluon dressing, the observed quantitative agreement is indicative 
of the level of sophistication reached by functional approaches in the past two decades; for a state-of-the-art survey, we refer to the recent reviews \cite{Huber:2018ned, Dupuis:2020fhh, Ferreira:2023fva}, as well as \cite{Huber:2020keu, Fu:2025hcm}. 

\begin{table}[t]
\vspace{.2cm}
\begingroup
\renewcommand{\arraystretch}{1.3}
\setlength{\tabcolsep}{6pt} 
\begin{tabular}{|c|c|c|c|c|}
	\hline
	$z_\textrm{c,ir}$  & 
    $z_\textrm{c,uv}$ & 
    $b_\textrm{uv}$ &
    $d_\textrm{ir}$[GeV$^{-2}$] &
    $d_\textrm{uv}$[GeV$^{-2}$] \\
	\hline
	$0.268$  & $0.616$ & $2.64$ & $1.14$ & $2.59$ \\
	  \hline
\end{tabular}
\caption{Parameters in \labelcref{eq:GhostFit} for the ghost propagator in Yang-Mills theory.\hspace*{\fill}}
\label{tab_GhostFitParameters}
\endgroup
\end{table}
%

\section{Functional approach to the confinement-deconfinement phase transition}
\label{app:FunConf-Deconf}

In this Appendix we briefly review 
the approach on confinement-deconfinement phase transition introduced in \cite{Braun:2007bx}.
At sufficiently high temperature, QCD enters the deconfined phase. 
In Yang-Mills theory, the system undergoes a second-order phase transition for two colors, and a first-order one for three colors or more. 
The underlying symmetry is the center symmetry of the gauge group, and the respective order parameter in Yang-Mills theory is the Polyakov loop. 
Loosely speaking, the Polyakov loop is related to the free energy of a static single quark, or rather, half of the energy of a static quark--anti-quark pair. 
In full QCD with dynamical fermions, the system exhibits a smooth crossover, as center symmetry is explicitly broken. 
In the following, in order to keep computations as simple as possible, 
we only consider Yang-Mills theory; for full 2+1 flavour QCD,  see e.g.,~\cite{Lu:2025cls}. 

In Yang-Mills theory,  
the free energy $F_{q {\bar q}}$ 
of a static quark--anti-quark pair at infinite distance is given by 
\begin{align}
\left\langle L[A_0]\right\rangle
\sim \exp \left[-\beta F_{q {\bar q}}/2 \right] \,, 
\label{eq:L-FreeEnergy}
\end{align}
where the traced Polyakov loop $L[A_0]$ is provided by \labelcref{eq:TracedP}. 

In the confined phase, $F_{q {\bar q}}$
diverges as the distance 
$r$ between the quark and the anti-quark increases. 
In turn, for large temperatures the system is becoming weakly-coupled, and the Polyakov loop can be expanded in powers of the gauge field around $A_0=0$. 
Hence, we find, 
\begin{align}
  \left\langle L[A_0]\right\rangle \ \left\{\begin{array}{rcl} = 0 & \quad  {\rm for}\quad & T<T_c\\[2ex] 
      > 0 & \quad {\rm for}\quad & T>T_c\end{array}\right. \,,
      \label{eq:exptrL}
\end{align}
where it is understood that the group direction that gives real and positive values for the Polyakov loop
has been singled out. 

The expectation value of the traced Polyakov loop \labelcref{eq:L-FreeEnergy} is an infinite order correlation function of the gauge field, which is challenging to compute within functional approaches that involve an expansion in correlation functions of powers of the fundamental field. 
In \cite{Braun:2007bx,Fister:2013bh} a related gauge-invariant order parameter has been defined. 
Moreover, in specific gauges this gauge-invariant order parameter has a linear relation to the expectation value of the gauge field in these gauges. 
The derivation of this relation uses the transformation properties of the Polyakov loop: it is the Wilson line winding around the time direction. In particular, the Polyakov loop 
lies in the gauge group, and can be represented as the exponential of an algebra element $\phi$, with 
 \begin{align}
 P(\vec x) = e^{2 \pi i \phi(\vec x)}\,,\qquad \phi(\vec x) \to U(\vec x) \phi(\vec x) U^\dagger(\vec x)\,, 
\label{eq:Polloop}
\end{align}
with $U(\vec x)\in SU(3)$ and periodic gauge transformations $U(\vec x)=U(t=0,\vec x)= U(t=\beta,\vec x)$. 
The covariant transformation properties of the algebra element $\phi(\vec x)$ in \labelcref{eq:Polloop} allow us to define the gauge-invariant order parameter, see \cite{Braun:2007bx},  
\begin{align} 
\langle \varphi(\vec x)\rangle\,,\qquad \varphi(\vec x)= \sum_n \nu_n\,\hat e_n\,, \quad \nu_n=\textrm{EV}[\phi(\vec x)]\,, 
\label{eq:EVPolloop}
\end{align} 
with the eigenmodes $\nu_n(\vec  x)$ of the matrix field $\phi$. 
The $\hat e_n$ in \labelcref{eq:EVPolloop} are the eigenvectors to the eigenvalue $\nu_n$. 
The eigenvalues $\varphi(\vec x)$ of the algebra field $\phi(\vec x)$ are gauge-invariant, due to the covariance of the algebra element. 
Moreover, their center-symmetric values are in one-to-one correspondence to a vanishing Polyakov loop: they both signal the center-symmetric confining phase. 

The mean field $\langle \varphi\rangle$ in \labelcref{eq:EVPolloop} can be readily computed from the effective potential of $A_0$, derived from the background DSE depicted in \Cref{fig:DSE-A}. 
To relate $\varphi$ to $A_0$, we use the freedom to take a specific gauge for the latter, as the background field effective action and hence the $A_0$-potential are background gauge-invariant: 
we use background fields $\bar A_0$ in the Polyakov gauge, where $A_0(x)=A_0^c(\vec x)$. Then, 
the temporal component is time-independent, and lies in the Cartan subalgebra,  indicated by the superscript $c$. 
Finally, take the background field that solves the equation of motion. Then we find 
\begin{align} 
\nu_n= \textrm{eigenvalues}\left(\frac{ \beta g_s A_0}{2 \pi}\right)\mod 1 \,,
\label{eq:EVvarphi}
\end{align}
and confinement is signalled by 
\begin{align} 
L[\langle \varphi\rangle ]=0\,, 
\label{eq:confPolphi}
\end{align}
where $\langle \varphi\rangle = \sum_n \langle \nu_n\rangle \hat e_n$. 
After the rotation in the Cartan subalgebra, the $\hat e_n$ for the non-vanishing eigenvalues are combinations of the Cartan matrices. 
In the simplest case of an SU(2) gauge theory, we have that 
$\nu_n\in(0, \pm \varphi_3)$. 
We also quote the eigenvalues for the physical SU(3) case,    
\begin{align}
    \nu_n\in \left( 0\,,\, 0\,,\, \pm \varphi_3\,,\, \pm \frac{\varphi_3\pm\sqrt{3}\,\varphi_8}{2}\right)\,, 
\label{eq:nu-varphi}
\end{align}
see e.g.,~\cite{Lu:2025cls}. 
  
The order parameter potential for $\varphi$ is readily computed in the background field approach to Yang-Mills theories~\cite{DeWitt:1967ub,tHooft:1971qjg,Honerkamp:1972fd,Kallosh:1974yh,Arefeva:1974jv,Kluberg-Stern:1974nmx,Weinberg:1980wa,Abbott:1980hw,Shore:1981mj,Abbott:1981ke,Abbott:1983zw}, where the covariant gauge-fixing is substituted by a gauge-covariant one: 
for that purpose, we split the gauge field
$A$ into two parts, 
a background (classical) component, $\bar A$,
and a fluctuating (quantum) field,  $a_\mu$, i.e., 
\begin{align} 
A_\mu=\bar A_\mu+a_\mu\,.
\label{eq:LinearSplit}
\end{align}
The respective gauge-fixed Yang-Mills action at finite temperature is given by 
\begin{align}
	S = \frac{1}{4}\int_x  F_{\mu\nu}^{a} F_{\mu\nu}^{a} + \int_x \left[ \bar{c}^a \bar{D}_{\mu} D_{\mu}^{ab} c^{b} + \frac{1}{2\xi} \left( \bar{D}_{\mu} a_{\mu} \right)^2 \right],
\label{eq:classical-action}
\end{align}
with ${D}_{\mu}(A) = \partial_{\mu} - \imag g_s \bar{A}_{\mu}$ and $\bar{D}_{\mu} = D_\mu(\bar A)$. 
The space-time integrals at finite temperature are given by 
\begin{align}
    \int_x = \int_0^\beta d x_0 \int\limits_{\mathbbm{R}} d^3 x\,,
\end{align}
with $\beta=1/T$. In the case of QCD, the 
kinetic and interaction terms 
associated with the quarks 
must be duly supplied to the 
action of \labelcref{eq:classical-action}. 

Then, the order parameter potential is a part of the effective action $\Gamma[\bar A, a,c,\bar c]$ for Yang-Mills theory, and $\Gamma[\bar A, a,c,\bar c, q,\bar q]$ for QCD, evaluated for given background gauge fields $\bar A$, and for vanishing fluctuation fields $a,c,\bar c, q,\bar q$; 
i.e., ~for Yang-Mills theory we have $ \Gamma[A]=\Gamma[\bar A=A, 0,0, 0]$. 
The background effective action $\Gamma[A]$ can be computed from its respective functional relation; 
in the present work we take the background DSE, see \cite{Fister:2013bh}. 
More precisely, we compute the temporal background gauge field DSE, which is depicted in \Cref{fig:DSE-A}. For constant backgrounds $A_0$, it defines the order parameter potential, see \Cref{sec:Veff}.

\section{RG-dependence of the fit \texorpdfstring{$Z_{A,\textrm{fit}}(p)$}{ZA}}
\label{app:ChangeMassgap}

The gluon dressing $Z_{A,\textrm{fit}}$ satisfies the RG-equation \labelcref{eq:RGZA}. 
Evidently, this implies a non-trivial $\mu$-dependence of the fitting parameters $z, c$. 
In this Appendix we use the UV-part $Z_{A,\textrm{uv}}$ in order to illustrate this $\mu$-dependence, as well as to discuss a reparametrisation, leading to $\mu$-independent fitting parameters. 

To that end, we restrict ourselves to RG-scales $\mu$ that are deep in the UV, such that  
\begin{align} 
Z_{A,\textrm{ir}}(\mu)\approx 0\,.
\label{eq:Zir=0}
\end{align}
\Cref{eq:Zir=0} holds true for the RG-scales used in the present work, namely with $\mu\gtrsim p_\textrm{in}^+$, and in particular for $\mu=4.3$\,GeV. This leads us to 
\begin{align}
    Z_{A,\textrm{fit}}(\mu) \approx Z_{A,\textrm{uv}}(\mu)\,.
\label{eq:RGConditionZA}
\end{align}
Then, a change of the RG-scale, $\mu\to \mu'$, while enforcing \labelcref{eq:RG-Conditions} at the new scale $\mu'$, leads to 
\begin{align} \nonumber 
z_\textrm{uv}\to &\,z_\textrm{uv}\left[1-c_\textrm{uv} \log \left(1+ c_+ x(\mu')\right)\right]^{\gamma}\,,\\[2ex]
c_\textrm{uv}\to &\,c_\textrm{uv}\left[
1-c_\textrm{uv} \log \left(1+ c_+ x(\mu)\right)\right]\,. 
\label{eq:TrafoNoTrafo}
\end{align} 
One readily sees that $Z_{A,\textrm{fit}}$ with  \labelcref{eq:TrafoNoTrafo} satisfies its RG-equation \labelcref{eq:RGZA} at one loop. 

\Cref{eq:TrafoNoTrafo} can also be used to define RG-invariant fitting parameters $\tilde c,\tilde z$. We quote the result,  
\begin{align}
Z_{A,\textrm{uv}}(x) = \tilde z_\textrm{uv}\left[1 + \tilde c_\textrm{uv}\log \frac{1+ c_+\,x}{1+c_+\,x(\mu)}\right]^\gamma\,, 
\label{eq:ZAuvAlt}
\end{align}
with 
\begin{align}\nonumber 
\tilde z_\textrm{uv} = & \,z_\textrm{uv} \left[1+c_\textrm{uv} \log \left(1+ c_+ x(\mu)\right) \right]^\gamma\,,\\[1ex]
\tilde c_\textrm{uv} =&\,\frac{c_\textrm{uv}}{1+c_\textrm{uv} \log \left(1+ c_+ x(\mu)\right) }\,, 
\end{align} 
with $\mu\, d (\tilde c,\tilde z)/d\mu=0$ (at one-loop). Note that a similar representation can also be achieved for the infrared part. 
However, this is considerably more difficult, reflecting the fact that the RG scale $\mu$ should rather be taken in the ultraviolet instead of the infrared regime. 

Moreover, \labelcref{eq:ZAuvAlt} gives us easy access to an analysis of the relation between the perturbative scale $\Lambda_\textrm{QCD}$ and the gluon mass gap $m_\textrm{gap}$, obtained from the Schwinger mechanism: 

\begin{figure}[ht]  \includegraphics[width=\linewidth]{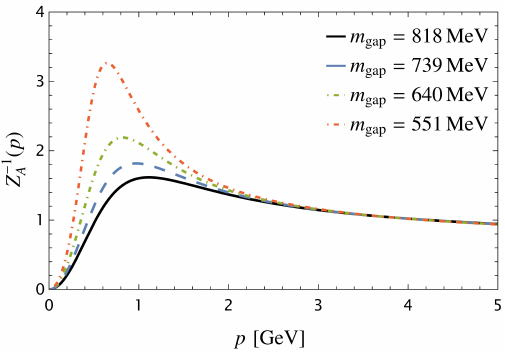}
  \caption{Gluon dressing for fixed $\Lambda_\textrm{QCD}$ and different gluon mass gaps $m_\textrm{gap}$. For $m_\textrm{gap} < 551$~MeV (dot-dashed red line) the propagator becomes complex for small momenta, signalling the onset of the Landau pole.  \hspace*{\fill}}
  \label{fig:ZAtildeLandauPole}
\end{figure}

In principle, the gluon mass gap may grow large, and hence $x(\mu)\to 0$. 
However, 
this is ruled out by the dynamics, 
as the longitudinal massless resonance triggering the Schwinger mechanism would not be generated by asymptotically small couplings. 

In turn, for $m_\textrm{gap}\to 0$, the fit \labelcref{eq:ZAuvAlt} runs into the perturbative Landau pole, see \Cref{fig:ZAtildeLandauPole}. 
Since it is precisely this Landau pole that the action of the Schwinger mechanism evades, $\Lambda_\textrm{QCD}$ defines an approximate infrared bound for the gluon mass gap. 
This concludes our discussion of the RG-scale dependence of $Z_{A,\textrm{fit}}$, and stresses again the natural identification of all scales in $Z_{A,\textrm{fit}}$, \labelcref{eq:GlobalFitGluon}, with the gluon mass gap. 

\bibliography{bib_master}

\end{document}